\documentclass{elsart}
\usepackage{epsfig}

\begin{document}

\begin{frontmatter}

\title{Application of the extended P+QQ force model
       to $N \approx Z$ $fp$ shell nuclei}

\author[Sawara]{M. Hasegawa}, \author[Higashi]{K. Kaneko} and
\author[Jonan]{S. Tazaki}
\address[Sawara]{Laboratory of Physics, Fukuoka Dental College,
 Fukuoka 814-0193, Japan}
\address[Higashi]{Department of Physics, Kyushu Sangyo University,
 Fukuoka 813-8503, Japan}
\address[Jonan]{Department of Applied Physics, Fukuoka University,
 Fukuoka 814-0180, Japan}

\begin{abstract}

    To study collective motion, the extended pairing plus $QQ$
 force model proposed recently is applied to $A$=46, 48  and 50 nuclei
 in the $fp$ shell region.
 Exact shell model calculations in the truncated model space
 $(f_{7/2},p_{3/2},p_{1/2})$ prove the usefulness of the interaction.
 The simple model with the pairing plus quadrupole pairing plus $QQ$
 force and $J$-independent isoscalar proton-neutron force reproduces
 unexpectedly well observed binding energies, energy levels of
 collective (yrast) states and reduced $E2$ transition probabilities
 in $^{46}$Ti, $^{46}$V, $^{48}$V, $^{48}$Cr, $^{50}$Cr and $^{50}$Mn.
 The correspondence between theory and experiment is almost comparable
 to that attained by the full $fp$ shell model calculations with
 realistic effective interactions.  Some predictions are made
 for energy levels and variations of $B(E2)$ in the yrast bands,
 in these nuclei.  Characteristics of the interaction are investigated
 by comparing with the realistic effective interactions.
 \\

\vspace*{3mm}
\leftline{PACS:  21.60.-n;21.10.Dr;21.10.Hw;21.10.Re}
\end{abstract}

\end{frontmatter}

\section{Introduction}

 Two approaches starting from different pictures are crossing
 in $fpg$ shell region: one is in progress along the pairing plus
 quadrupole ($P_0+QQ$) force model
 \cite{Kisslinger,Baranger,Kishimoto,Bohr,Hara1,Hara2} and the other
 is the shell model approach using various effective interactions
 \cite{Kuo,Gloeckner,Serduke,Gross,Poves,Blom,Ji,Richter,Sin}.
 With the advancement of computer, the shell model approach has been
 extending the scale of calculation and the sphere of study
 \cite{Caurier,Rich2,Rudolph,Zamick,Herndl,Martinez,Martinez2,Poves2}.
 The full $fp$ shell model calculations with the KB3 interaction
 \cite{Caurier,Martinez,Martinez2,Poves2} have successfully explained
 the microscopic structure of the $f_{7/2}$ shell nuclei,
 and are pushing the front of study to $A \approx 52$.
 On the other hand, the $P_0+QQ$ force written in the isospin
 invariant form seems to qualitatively explain some properties of
 nuclei in the $fp$ shell region \cite{Langanke,Kaneko,Kaneko2}.
 One could expect a relationship between the two approaches and
 between the two types of interactions.  In fact, it was shown
 \cite{Dufour} that an important part of the realistic interactions
 is approximated by the $P_0+QQ$ force.   The $P_0+QQ$ force
 accompanied by the quadrupole pairing ($P_2$) force has succeeded
 in explaining the backbending mechanism of $^{48}$Cr \cite{Hara}.
 We note here that including both of $T=0$ and $T=1$ proton-neutron
 ($p$-$n$) interactions in the $P_0+QQ$ force is recommended
 \cite{Kaneko}.
   
  In a recent paper \cite{Hasegawa}, we have shown that the $P_0+QQ$
 force can be extended so as to get near to realistic effective
 interactions.  The extended interaction is composed of the four
 typical forms of interactions, i.e., the isospin invariant
 $P_0+P_2+QQ$ force and $J$-independent isoscalar $p$-$n$ force
 ($V^{\tau =0}_{\pi \nu}$).  This form very harmonizes with the design
 of Ref. \cite{Dufour}. The calculations \cite{Hasegawa,Kaneko3}
 showed that the isoscalar $p$-$n$ force $V^{\tau =0}_{\pi \nu}$ plays
 a decisive role in reproducing the gross nuclear properties such as
 the binding energy, symmetry energy, etc., and that the
 $P_0+P_2+QQ+V^{\tau =0}_{\pi \nu}$ interaction describes considerably
 well the $f_{7/2}$ and g$_{9/2}$ shell nuclei in which a subshell
 dominates.  The usefulness of the model, however, was studied
 within the single $j$ shell model calculations there.
 The $P_0+QQ$ force model originally aims to describe nuclear
 collective motion.  The next crucial step is, therefore, to examine
 whether the extended $P_0+QQ$ force model is quantitatively applicable
 to real situations having many $j$ shells and is capable of describing
 the collective motion or not.
 This examination will show to what extent the picture of the $P_0+QQ$
 force model persists by a modification in various regions.
 We have confirmed that our interaction can describe the global
 features of $p$-$n$ interactions and symmetry energy in another paper
 \cite{Kaneko4}.
 The purpose of this paper is to investigate in detail the
 applicability of the $P_0+P_2+QQ+V^{\tau =0}_{\pi \nu}$ interaction
 model by a realistic treatment of the $f_{7/2}$ shell nuclei.  
   
  We carry out exact shell model calculations in even-$A$ nuclei with
 $A=44-50$ in order to see products of the interaction without any
 disturbance due to approximate treatment.  This makes it easy to
 compare our interaction with the realistic effective interactions.
 We must determine the four force strengths when dealing with the
 $P_0+P_2+QQ+V^{\tau =0}_{\pi \nu}$ interaction.
 Unfortunately, the authors do not have a fast computer code enough
 for searching appropriate force strengths in the full $fp$ shell.
 We therefore adopt the configuration space ($f_{7/2}$, $p_{3/2}$,
 $p_{1/2}$) as a model space.  Calculations prove the usefulness of
 this model space especially for our interaction including the
 $V^{\tau =0}_{\pi \nu}$ force (the reason will be explained in Sec.
 IV).  Our simple model with only four force strengths in the small
 configuration space gives unexpectedly good results in $A$=46, 48
 and 50 nuclei with $N \approx Z$, which almost match the full $fp$
 shell model calculations with the KB3 interaction
 \cite{Caurier,Martinez,Martinez2,Poves2},
 so that we can discuss properties of these nuclei in our model
 and give information unreported previously.
   
  This paper is organized as follows.  In Sec. II, we briefly review
 the model.  The results of calculations on the binding energies,
 energy levels and electric transition probabilities $B(E2)$ are
 shown for $^{46}$Ti, $^{46}$V, $^{48}$V, $^{48}$Cr, $^{50}$Cr and
 $^{50}$Mn in Sec. III.  The dependence of the results on the model,
 the relation of our interaction to the realistic effective
 interactions and properties of the collective yrast bands are
 discussed in Sec. IV.  Concluding remarks are given in Sec. V.

\section{The model}

 We have proposed the following isospin-invariant Hamiltonian composed
 of the four forces, $P_0$, $P_2$, $QQ$ and $V_{\pi \nu}^{\tau =0}$,
 and have discussed the basic properties of it in Ref. \cite{Hasegawa}:
\begin{eqnarray}
 & & H = H_{\rm sp} + V(P_0) + V(P_2) + V(QQ)
  + V^{\tau =0}_{\pi \nu}, \label{eq:1} \\
 & {} & H_{\rm sp}= \sum_a \epsilon_a ( \hat{n}_{a \pi}
  + \hat{n}_{a \nu} ),                               \label{eq:2} \\
 & {} & V(P_J) = - {1 \over 2} g_{J} \sum_{M \kappa}
  \sum_{a \leq b} \sum_{c \leq d} p_J(ab) p_J(cd)
  A^\dagger_{JM1\kappa}(ab) A_{JM1\kappa}(cd),       \label{eq:3} \\
 & {} & V(QQ) = -{1 \over 2} \chi^\prime \sum_M \sum_{ac \rho}
  \sum_{bd \rho^\prime} q^\prime(ac) q^\prime(bd)
  : B^\dagger_{2M \rho}(ac) B_{2M \rho^\prime}(bd):  \label{eq:4} \\
 & {} & V^{\tau =0}_{\pi \nu}= - k^0 \sum_{a \leq b} \sum_{JM}
    A^\dagger_{JM00}(ab) A_{JM00}(ab),               \label{eq:5}
\end{eqnarray}
with
\begin{eqnarray}
 {} & A^{\dagger}_{JM \tau \kappa}(ab) = { [ c^\dagger_a
 c^\dagger_b ]_{JM \tau \kappa} / \sqrt{1 + \delta_{ab} } },
 \quad
 B^\dagger_{JM \rho}(ac) = [ c^\dagger_{a \rho} c_{c \rho} ]_{JM},
                                                      \label{eq:6} \\
 {} & p_{0}(ab)=\sqrt{2j_a+1} \delta_{ab} , \quad
 p_{2}(ac)=q^\prime(ac) = (a\|r^2Y_2/b_0^{2}\|c) /\sqrt{5}.
                                                      \label{eq:7}
\end{eqnarray}
 Here $\hat{n}_{a \pi}$ and $\hat{n}_{a \nu}$ are the proton- and
 neutron-number operators for a single-particle orbit $a$,
 the symbol {: :} in $V(QQ)$ means the normal order product of
 four nucleon operators, $\rho$ denotes the $z$-components of isospin
 $\pm\frac{1}{2}$ and $b_0$ is the harmonic-oscillator range parameter.
 This Hamiltonian has only the four force parameters, $g_{0}$, $g_{2}$,
 $\chi^\prime$, and $k^{0}$, in addition to the single-particle
 energies $\epsilon_a$. We call the $P_0+P_2+QQ+V^{\tau =0}_{\pi \nu}$
 interaction ``functional effective interaction (FEI)" in this paper.

 As shown in Ref. \cite{Hasegawa}, the $J$-independent isoscalar
 $p$-$n$ force $V_{\pi \nu}^{\tau =0}$ is reduced to 
\begin{equation}
  V^{\tau=0}_{\pi \nu} = - {1 \over 2} k^0
   \{ {{\hat n} \over 2}({{\hat n} \over 2}
   +1) - {\hat {T^2} } \} ,      \label{eq:8}
\end{equation}
where the operator $\hat n$ stands for the total number of the valence
 nucleons and $\hat {T}$ for the total isospin, i.e., 
 $ \hat n=\hat n_p+\hat n_n
 = \sum_a ( {\hat n_{a\pi}} + {\hat n_{a\nu}} ) $ 
 and $\hat {T}=\sum_a \hat {T_a}$.
 For the states with the isospin $T=(n_n-n_p)/2$,
 this force can be rewritten as
\begin{equation}
 \langle V^{\tau=0}_{\pi \nu} \rangle
  = - {1 \over 2} k^0 n_p (n_n +1).   \label{eq:9}
\end{equation}
 In our isospin invariant Hamiltonian, two conjugate nuclei with
 $T=(n_n-n_p)/2=(n_p-n_n)/2$ are equivalent to each other, and the
 $T=1$ states in odd-odd nuclei with $N=Z=A/2$ are equivalent to
 those in even-even nuclei with $Z=A/2-1$ and $N=A/2+1$.
 The $p$-$n$ force $V_{\pi \nu}^{\tau =0}$ is indispensable for the
 binding energy but does not change the wave functions determined
 by the $P_0+P_2+QQ$ force.  The $V_{\pi \nu}^{\tau =0}$ force is
 the main origin of the symmetry energy \cite{Kaneko3} and brings
 a shift to the energy space between the states with different
 isospins in a nucleus.
 The shell model Hamiltonian is originated in the Hartree-Fock (HF)
 picture and the single-particle energies $\epsilon_a$ stand for the
 single-particle mean field determined by the HF theory.  The $p$-$n$
 force $V_{\pi \nu}^{\tau =0}$ in the simple expression (\ref{eq:9})
 suggests that $V_{\pi \nu}^{\tau =0}$ is probably related to the HF
 variation and may be an average term by which the mean field is
 accompanied.  If we extract $V_{\pi \nu}^{\tau =0}$, residual
 interaction in the mean field comes from the $P_0$, $P_2$ and $QQ$
 forces in our model.  This interpretation is similar to that of
 Dufour and Zuker \cite{Dufour}. The residual $P_0+P_2+QQ$
 interaction can be understood based on the Bohr-Mottelson picture
 \cite{Bohr}. This problem how to explain the phenomenologically
 introduced $p$-$n$ force $V_{\pi \nu}^{\tau =0}$ should be
 discussed further.

 The Hamiltonian (\ref{eq:1}-\ref{eq:5}) describes the energy of
 valence nucleons outside the doubly-closed-shell core ($^{40}$Ca)
 excluding the Coulomb energy.  The corresponding experimental energy
 is evaluated as
\begin{equation}
 W_0(Z,N) = B(Z,N) - B(^{40}{\mathrm{Ca}}) - \lambda (A-40)
 - \Delta E_{\mathrm{C}}(n_p,n_n), \label{eq:10}
\end{equation}
where $B(Z,N)$ is the nuclear binding energy and $\lambda$ is the
 base level of single-particle energies.  We fix $\lambda=-8.364$ MeV
 so that $W_0(^{41}\mathrm{Ca})=W_0(^{40}\mathrm{Ca})=0.0$ and the
 lowest single-particle energy $\epsilon_{7/2}$ becomes zero,
 and evaluate the Coulomb energy correction 
 $\Delta E_{\mathrm{C}}(n_p,n_n)$ by the function \cite{Pasquini}
\begin{equation}
 \Delta E_{\mathrm{C}}(n_p,n_n) = 7.279 \, n_p + 0.15 \, n_p
  (n_p -1) -0.065 \, n_p n_n . \label{eq:11}
\end{equation}
We shall compare calculated ground-state energies with the
 experimental energies $W_0(Z,N)$.

   We adopt the model space $(f_{7/2}, p_{3/2}, p_{1/2})$ as
 mentioned in Introduction.
 Results of calculations depend on the single-particle energies
 $\epsilon_a$ and the four force parameters $g_0, g_2, \chi^\prime$
 and $k^0$.   We take $\epsilon_a$ from the experimental spectrum
 of $^{41}$Ca (we fix $\epsilon_{7/2}=0$ mentioned above) as follows:
\begin{equation}
\epsilon_{7/2}=0.0, \quad  \epsilon_{3/2}=1.94, \quad
 \epsilon_{1/2}=3.61 \quad   \mbox{ in MeV}.  \label{eq:12}
\end{equation}
We employed fixed numbers for $g_0, g_2$ and $\chi^\prime$ in the
 single $j$ model in Refs. \cite{Hasegawa,Kaneko3}.  Calculations,
 however, recommend using $A$-dependent parameters for the $P_0$,
 $P_2$ and $QQ$ forces in the present many $j$ shell case as the
 ordinary treatment of these forces.  We put the same $1/A$ dependence
 on $k^0$ as in Refs. \cite{Hasegawa,Kaneko3}.  A rough parameter
 search in $A=46-50$ nuclei leads us to the values
\begin{eqnarray}
 & { } & g_0=0.48 (42/A), \quad g_2=0.36 (42/A)^{5/3},
  \nonumber \\
 & { } & \chi^\prime=0.30 (42/A)^{5/3},  \quad k^0=2.23 (42/A),
  \quad \mbox{ in MeV}. \label{eq:13}
\end{eqnarray}
We do not readjust the seven parameters in each nucleus.

   The present calculations are considerably realistic and are
 expected to provide good information about wave functions.  To test
 the wave functions, we calculate the reduced quadrupole transition
 probabilities $B(E2)$ and compare them with observed values.
 Following Caurier \textit{et al.} \cite{Caurier}, we also calculate
 the intrinsic quadrupole moment $Q_0$ by the equations
\begin{eqnarray}
   & Q_0={{(J+1)(2J+3)} \over {3K^2-J(J+1)}} Q_{spec}(J) 
   \quad \mbox{ for } K \not= 1,    \label{eq:14} \\
   & B(E2:J_i \rightarrow J_f)={5 \over 16\pi} e^2
    \langle J_iK20|J_f K \rangle^2 Q^2_0,    \label{eq:15}
   \end{eqnarray}
where $Q_{spec}(J)$ is the spectroscopic quadrupole moment
 $\sqrt{16\pi /5} \langle JJ|Q^\dagger_{20}|JJ \rangle$ with
 $Q^\dagger_{2M\rho}=e_\rho b_0^2 \sum_{ac} q^\prime(ac)
  [ c^\dagger_{a \rho} c_{c \rho} ]_{2M}$
 and $K$ is the projection of the total spin on the intrinsic axis.
 We use the harmonic-oscillator range parameter $b_0=1.01A^{1/6}$ fm,
 and effective charges of 1.5$e$ for protons and 0.5$e$ for neutrons
 in the calculations of $B(E2)$ and $Q_0$.  These values are the same
 as those used in Refs.
 \cite{Caurier,Zamick,Martinez,Martinez2,Poves2}.
 We denote the intrinsic quadrupole moment calculated from Eq.
 (\ref{eq:14}) by $Q^{(s)}_0$ and that calculated from Eq.
 (\ref{eq:15}) by $Q^{(t)}_0$ following Mart\'{\i}nez-Pinedo
 \textit{et al.} \cite{Martinez2}.

\section{The results of calculations}

  We have carried out exact shell model calculations with the model
 shown in Sec. II in $A$=46, 48 and 50 systems. The ground-state
 energies obtained are listed in Table \ref{table1}, where the
 experimental energies $W_0$ are calculated from mass excesses
 \cite{ToI} using Eqs. (\ref{eq:10}) and (\ref{eq:11}).  Our simple
 interaction $P_0+P_2+QQ+V^{\tau=0}_{\pi \nu}$ reproduces the
 experimental energies well.  The agreement is excellent for the
 $N=Z$ nuclei, $^{46}$V, $^{48}$Cr and $^{50}$Mn.  Note that our
 model systematically reproduces the energies of different-number
 nuclei within a single set of parameters.  The results confirm the
 essential role of the $J$-independent isoscalar $p$-$n$ force
 $V^{\tau=0}_{\pi \nu}$ in the binding energies of $N \approx Z$
 nuclei which is stressed in our previous papers
 \cite{Hasegawa,Kaneko3}.  It is certain that an important part of
 nucleon-nucleon interactions can be written as
 $V^{\tau=0}_{\pi \nu}$.

\begin{footnotesize}
\begin{table}[h]
\caption{Ground-state energies of $A=46, 48, 50$ nuclei.}
\label{table1}
\begin{center}
\begin{tabular}{ccccccc}\hline
   & $^{46}$Ti($^{46}$V,0$^+$) & $^{48}$Ti & $^{48}$V(4$^+$) &
   $^{48}$Cr & $^{50}$V(6$^+$) & $^{50}$Cr($^{50}$Mn,0$^+$) \\ \hline
  $W_0$ & -20.28 & -23.82 & -26.71  & -32.38 & -30.48 & -38.71 \\
  Calc. & -20.17 & -23.44 & -26.79  & -32.38 & -30.00 & -38.72 \\
   \hline
\end{tabular}
\end{center}
\end{table}
\end{footnotesize}

   The ${P_0+P_2+QQ+V^{\tau=0}_{\pi \nu}}$ interaction, however, 
 cannot give enough binding energies as $N$ separates from $Z$.
 Especially, the disagreement becomes large when the number of valence
 protons or neutrons is 0 and 1.  This is in contrast to that the KB3
 interaction overbinds the $A$=48 nuclei about 0.78 MeV but provides
 very good relative binding energies of all the $A$=48 nuclei
 \cite{Caurier}. We shall discuss the difference of our interaction
 from the realistic interactions in the next section.

  In the following subsections, we show calculated excitation energies
 and electric quadrupole properties in $^{46}$Ti, $^{46}$V, $^{48}$V,
 $^{48}$Cr, $^{50}$Cr and $^{50}$Mn with $Z, N \geq 42$ where the 
 ${P_0 + P_2 + QQ + V^{\tau=0}_{\pi \nu}}$ interaction works well.  
 Only positive-parity states are considered in our model space.
 It should be remembered here that the energy spectra and wave
 functions are determined by the ${P_0+P_2+QQ }$ force in our model,
 and the ${V^{\tau=0}_{\pi \nu}}$ force causes only an energy shift
 depending on the valence nucleon number and isospin.

\subsection{$^{46}$Ti}

\begin{figure}[h]
\begin{center}
    \epsfig{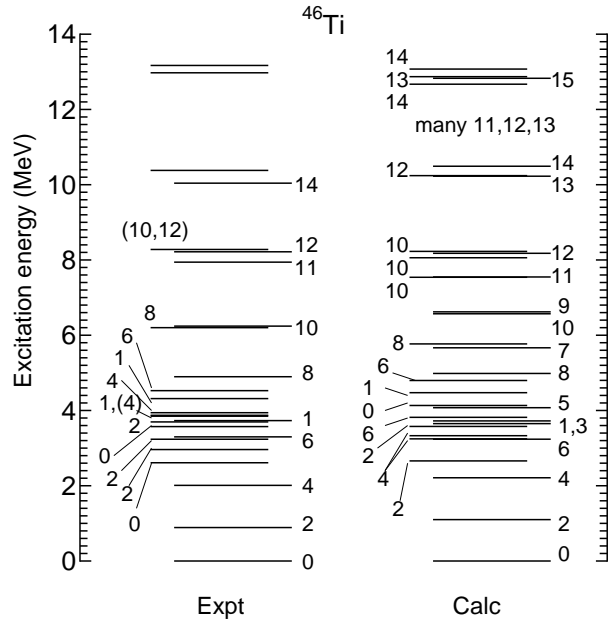}
\caption{Calculated and observed energy levels of $^{46}$Ti.}
\label{fig1}
\end{center}
\end{figure}

  Calculated excitation energies and $E2$ transition probabilities in
 $^{46}$Ti are compared with observed ones \cite{ToI,Peker,Cameron2}
 in Fig. \ref{fig1} and Table \ref{table2}.  The yrast levels
 (the lowest-energy state of each spin) are satisfactorily reproduced
 in spite of the simple model.  We can regard the $0_1^+$, $2_1^+$,
 $4_1^+$ ... $14_1^+$ states as the ground-state band.
 They are connected by the large $E2$ transition probabilities,
 especially up to the $10_1^+$ state.
  The observed $B(E2:J \rightarrow J-2)$ values in the ground-state
 band are considerably well reproduced by our model though the model
 space does not include the $f_{5/2}$ orbit.
  The spectroscopic quadrupole moments ($Q_{spec}$)
 of the $0_1^+$, $2_1^+$..... $14_1^+$ states have a minus sign and
 their values are in the range 19-26 $e$ fm$^2$.  The intrinsic
 quadrupole moments $Q_0^{(s)}$ and $Q_0^{(t)}$ are large and roughly
 constant up to the 10$_1^+$ state, which tells us that these states
 are collective and have the same nature.
 The observed energy levels of the ground-state band show the
 backbending at the $10_1^+$ state. The calculated energy levels also
 show a similar tendency, though the position of the $10_1^+$ state is
 not very good. The calculated intrinsic quadrupole moment $Q_0^{(t)}$
 suggests different structure at least above the $10_1^+$ state.

 From the good reproduction of the $1_1^+$ and $11_1^+$ levels as well
 as the ground-state band, the calculated energy levels of the $J$=odd
 yrast states probably provide a good prediction for their positions.
 This will be tested in future experiments.

   In Fig. \ref{fig1}, we also compare calculated low energy levels
 below 4 MeV and a couple of levels above it with observed ones,
 where we omit 4 levels assigned as $J^\pi = 2^+$ and 7 other levels
 with indefinite spin and parity \cite{ToI} from the experimental
 column in order to avoid overcrowding.  We can suppose that some of
 the observed states including the $0_2^+$ state at 2.661 MeV are
 core-excited states (probably 8-particle-2-hole ones) on the analogy
 of the core-excited states in $^{44}$Ti.  
 The calculated energy levels of non-collective states lie at a good
 position roughly speaking and the level density looks good if we
 exclude the core-excited states.

  The state at 10.040 MeV, which was assigned as $J$=12 before
 \cite{Cameron1}, was recently identified as $J$=14 \cite{Cameron2},
 because the transition to $11_1^+$ was not detected.
  The calculated $B(E2)$ values, $B(E2:14_1 \rightarrow 12_1)$=26,
 $B(E2:12_2 \rightarrow 12_1)$=3.3 and
 $B(E2:13_1 \rightarrow 12_1)$=0.1 in $e^2$ fm$^4$, suggest that
 the two levels at 10.040 MeV and 10.380 MeV are probably $14_1^+$
 and $12_2^+$, though their order is inverse in our prediction.
 The present calculation also provides candidates for the two observed
 levels at 12.974 MeV and 13.169 MeV. They are possibly the $14_2^+$
 and $15_1^+$ states from the calculated $B(E2)$ values.

\begin{table}[b]
\caption{Electric quadrupole properties of the yrast and other states
 in $^{46}$Ti. The observed $B(E2)$ values are taken from Refs. [30]
 (Expt.1) and [31] (Expt.2).}
\label{table2}
\begin{center}
\begin{tabular}{rcccccc} \hline
$J_n \rightarrow J^\prime_m$ &
  \multicolumn{3}{c}{$B(E2:J_n - J^\prime_m)$ in $e^2$ fm$^4$} 
 & \multicolumn{3}{c}{calculated $Q_{spec}$, $Q_0$ in $e$ fm$^2$} \\
 & Expt.1 & Expt.2 & Calc. & $Q_{spec}(J_n)$ & $Q^{(s)}_0$
 & $Q^{(t)}_0$  \\ \hline
 $2_1 \rightarrow 0_1$ & 197$\pm$32 &  & 119 & -18.9 & 66  & 77 \\
 $4_1 \rightarrow 2_1$ & 192$\pm$32 &  & 159 & -24.5 & 67  & 75 \\
 $6_1 \rightarrow 4_1$ & 158$\pm$28 &  & 141 & -21.3 & 53  & 67 \\
 $8_1 \rightarrow 6_1$ & 119$\pm$38,
  & 62$\pm$30 & 126 & -23.7 & 56 & 62 \\
 $8_2 \rightarrow 6_1$ &  &  & 0.03 & -11.1 & 27 &  \\
     $\rightarrow 8_1$ &  &  & 8.8  &      &     &  \\
$10_1 \rightarrow 8_1$ & 152$\pm$15,
  & 77$\pm$36 & 104 & -26.3 & 61 & 56 \\  \hline
$10_2 \rightarrow 10_1$ &  &  & 3.8 & -12.1 & 28 &  \\
     $\rightarrow 8_1$  &  &  & 3.7 &      &     &  \\
     $\rightarrow 8_2$  &  &  & 46  &      &     &  \\
$12_1 \rightarrow 10_1$ &  25$\pm$10, & 22$\pm$7 & 53 & -22.3
       & 50 & 39 \\
$12_2 \rightarrow 10_1$ &  &  & 17  & -18.9 & 43 &  \\
     $\rightarrow 11_1$ &  &  & 7.8 &      &     &  \\
     $\rightarrow 12_1$ &  &  & 3.3 &      &     &  \\
     $\rightarrow 10_2$ &  &  & 0.3 &      &     &  \\
$13_1 \rightarrow 12_1$ &  &  & 0.1 &      &     &  \\
$14_1 \rightarrow 12_1$ &  46$\pm$30, &   $>$44  & 26 & -22.8
       & 51 & 27 \\
     $\rightarrow 13_1$ &  &  & 9.5 &      &     &  \\
$15_1 \rightarrow 14_1$ &  &  & 2.6 &      &     &  \\
     $\rightarrow 14_2$ &  &  & 3.0 &      &     &  \\
$14_2 \rightarrow 14_1$ &  &  & 6.2 & -23.1 & 51 &  \\
     $\rightarrow 12_2$ &  &  & 12  &      &     &  \\ \hline
\end{tabular}
\end{center}
\end{table}

\subsection{$^{46}$V}

\begin{figure}[h]
\begin{center}
    \epsfig{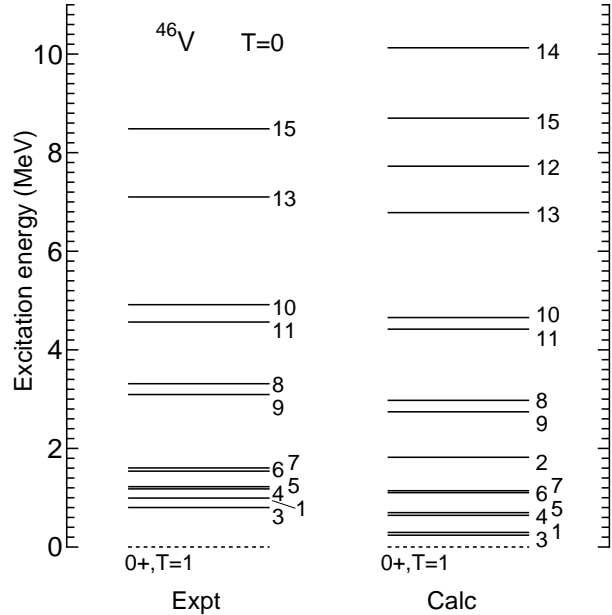}
\caption{Calculated and observed energy levels of $^{46}$V.
         Only the yrast states with $T=0$ are shown.}
\label{fig2}
\end{center}
\end{figure}

  Recently, the $N=Z$ odd-odd nucleus $^{46}_{23}V_{23}$ has been
 studied by elaborate experiments \cite{Friessner,Lenzi2}.  Let us
 compare calculated energy levels of the $T=0$ yrast states with
 observed ones in Fig. \ref{fig2}.  Our interaction somewhat overbinds
 the $T=0$ low-spin states as the realistic effective interactions do
 \cite{Friessner,Lenzi2}.
  It is still notable that such a simple force $V^{\tau=0}_{\pi \nu}$
 improves so well both the binding energy and relative energy between
 the $T=0$ and $T=1$ states.  If we regard the $T=0$ yrast states
 $3_1^+$, $4_1^+$, $5_1^+$ ... $15_1^+$ as a collective band based on
 the $3_1^+$ state, the theoretical pattern of excitation is very
 similar to the observed pattern.  The order of spin in the band is
 correctly reproduced.  The correspondence of the excitation energies
 measured from the $3_1^+$ level between theory and experiment is good.

  The calculated $B(E2)$ values listed in Table \ref{table3} support
 that the yrast states $3_1^+$, $4_1^+$, $5_1^+$, $6_1^+$ and $7_1^+$
 in order of spin $J$ are members of a collective band, because they
 are connected by large $B(E2:J \rightarrow J-1)$.  The spectroscopic
 quadrupole moment $Q_{spec}$ shows a rapid change as $J$ increases
 till $8_1^+$ , and then becomes nearly constant
 (-20 to -25 $e$ fm$^2$). Above $7^+_1$, the pair states with even
 spin $2J$ and odd spin $2J+1$ reverse their order in energy and
 the two series of states with odd $J$ and even $J$ connected by large
 $B(E2: J \rightarrow J-2)$ stand out.
  We can guess a structure change in the band between $7_1^+$ and
 $9_1^+$. Is it possible to regard the two series, $9_1^+$, $11_1^+$
 .... $15_1^+$ and $8_1^+$, $10_1^+$ ....$14_1^+$, as two quasi-bands?
 The $\gamma$ transitions observed in Ref. \cite{Lenzi2} seem to
 display this feature.
 
\begin{table}
\caption{Calculated $B(E2)$ (in $e^2$ fm$^4$) and $Q_{spec}$
 (in $e$ fm$^2$) with respect to the $T=0$ yrast states in $^{46}$V.}
\label{table3}
\begin{center}
\begin{tabular}{rcc|rcc}\hline
$J \rightarrow J^\prime$ & $B(E2)$ & Q$_{spec}$ & $J
 \rightarrow J^\prime$ & $B(E2)$ & $Q_{spec}$   \\ \hline
3                 &     &  33.2 & 10 $\rightarrow$  8 &  91 & -25.2 \\
4 $\rightarrow$ 3 & 249 &  12.0 &    $\rightarrow$  9 &  24 &       \\
5 $\rightarrow$ 3 &  60 &  -4.1 &    $\rightarrow$ 11 &  31 &       \\
  $\rightarrow$ 4 & 207	&       & 11 $\rightarrow$  9 & 129 & -21.9 \\
6 $\rightarrow$ 4 &  85 & -10.6 & 12 $\rightarrow$ 10 &  71 & -21.2 \\
  $\rightarrow$ 5 & 165 &       &    $\rightarrow$ 11 &  7.2&       \\
7 $\rightarrow$ 5 &  46 & -14.8 &    $\rightarrow$ 13 &  22 &       \\
  $\rightarrow$ 6 & 122 &       & 13 $\rightarrow$ 11 &  99 & -23.3 \\
8 $\rightarrow$ 6 & 111 & -21.5 & 14 $\rightarrow$ 12 &  33 & -24.0 \\
  $\rightarrow$ 7 &  68 &       &    $\rightarrow$ 13 &  7.6&       \\
9 $\rightarrow$ 7 & 142 & -20.3 &    $\rightarrow$ 15 &  44 &       \\
8 $\rightarrow$ 9 &  75 &       & 15 $\rightarrow$ 13 &  65 & -22.0 \\
  \hline
\end{tabular}
\end{center}
\end{table}

\begin{figure}
\begin{center}
    \epsfig{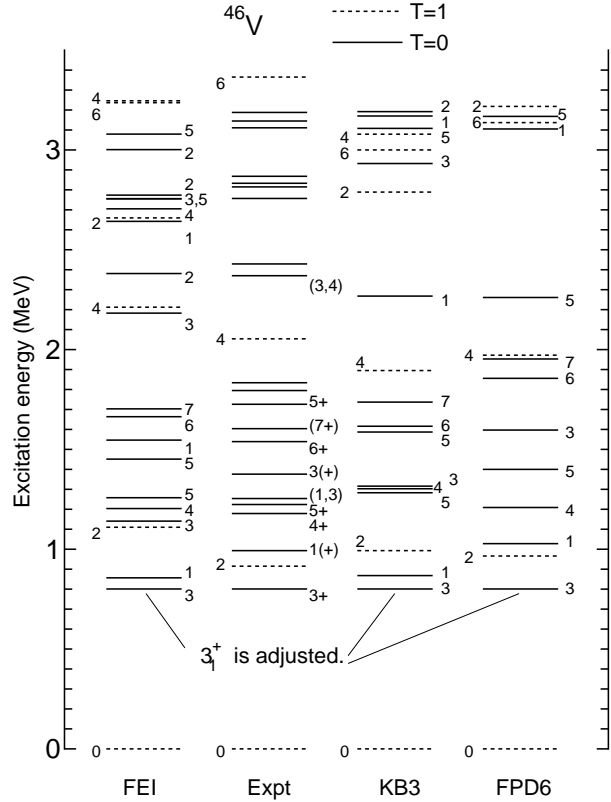}
\caption{Energy levels obtained using various effective interactions
        in $^{46}$V, compared with observed ones.  Observed levels
        with unknown spin and parity are also shown.}
\label{fig3}
\end{center}
\end{figure}

   In Ref. \cite{Friessner}, the low-lying states of $^{46}$V were
 examined by the full $fp$ shell model calculations with the realistic
 effective interactions, KB3 \cite{Caurier} and FPD6 \cite{Richter}.
 We compare the result obtained using our interaction FEI with the
 results of Ref. \cite{Friessner}, in Fig. \ref{fig3}.
 In this figure, all the energy levels below 3.25 MeV including
 unknown-parity levels but excluding the negative-parity ones are
 shown ($6_1^+$ and $4_2^+$ with $T=1$ are added).  The calculated
 $T=0$ levels are shifted so that $3_1^+$ is situated at the observed
 position 0.801 MeV as done in Ref. \cite{Friessner}.  The shift 0.56
 MeV for the FEI is a little larger than 0.436 MeV for the KB3 and
 0.498 MeV for the FPD6.  For the $T=0$ states, the FEI result seems
 to be better than the FPD6 one and considerably match the KB3 one.
 The FEI result for the $T=1$ levels is not very bad as compared with
 the KB3 and FPD6 results.  The non-collective states except the yrast
 states obtained by the FEI have a tendency to go down (this tendency
 is observed in other systems).  The number of the energy levels
 obtained by the FEI may be too many below 3.25 MeV but is smaller
 than that of the observed levels. The levels shown in the column Expt
 possibly include core-excited states \cite{Friessner}.  The 8p-2h
 states with positive parity in addition to the negative-parity states
 could be at low energy, because the $(fp)^8$ configurations
 corresponding to $^{48}$Cr gain very large binding energy due to
 $\alpha$-like four nucleon correlations discussed in the next section.
 The observed energy levels with $T=0$ seem to be in a middle
 situation between the KB3 and FEI results.

\subsection{$^{48}$Cr}

\begin{figure}[t]
\begin{center}
    \epsfig{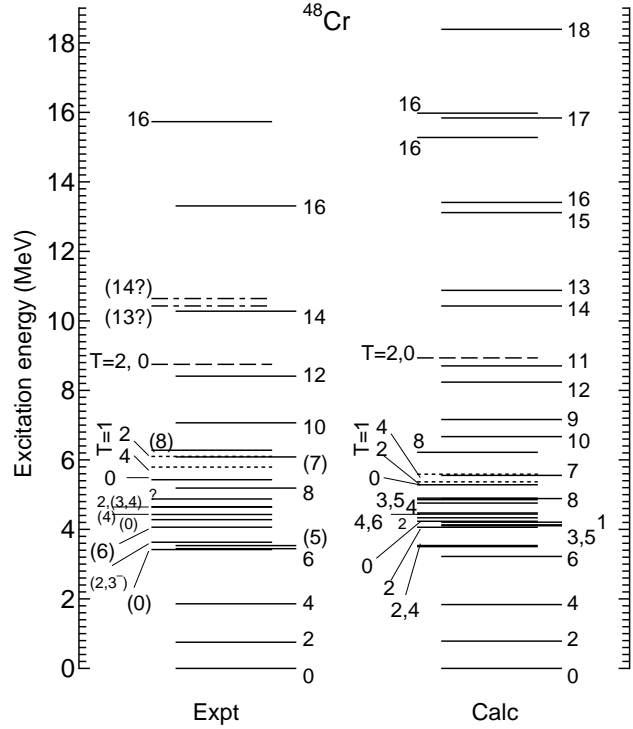}
\caption{Calculated and observed energy levels of $^{48}$Cr.}
\label{fig4}
\end{center}
\end{figure}

   The $^{48}$Cr nucleus has been attracting notice to the rotational
 band and backbending.  In Fig.\ref{fig4}, we illustrate calculated and
 observed energy levels \cite{ToI,Brandolini,Cameron3,Cameron4}.  
 The observed yrast levels, all levels with $J \leq 6$ below 5 MeV and
 some others above 5 MeV are shown in the column Expt and 
 corresponding levels obtained by the present model are shown in the 
 column Calc.  In Table \ref{table4}, we compare electric 
 quadrupole properties between theory and experiment \cite{Brandolini}
 (the observed value of $B(E2:2_1^+ \rightarrow 0_1^+)$ is from Ref.
 \cite{Burrows1}) and also show the full $fp$ shell model result of
 Caurier \textit{et al.} \cite{Caurier} for comparison.

   Our model reproduces satisfactorily well the observed levels of the
 ground-state band and the backbending at the $12_1^+$ level, though
 there are slight deviations for the $6_1^+$, $8_1^+$ and $10_1^+$
 states. The agreement is nearly comparable to that of the full $fp$
 shell model calculation with the KB3 interaction \cite{Caurier}.
 The calculated $E2$ transition probabilities $B(E2:J \rightarrow J-2)$
 in the ground-state band, which are also comparable to those of 
 Caurier \textit{et al.}, correspond well with the observed very large
 $B(E2)$ values in spite of the lack of the $f_{5/2}$ orbit. This
 suggests that our interaction including the $QQ$ force is in harmony
 with the collective motion of the ground-state band and yields
 the large $E2$ transition probabilities.  If we add the $f_{5/2}$
 orbit to the present model, a smaller effective charge can probably
 reproduce the observed $B(E2)$ values.  The spectroscopic quadrupole
 moment $Q_{spec}$ indicates the structure change of the ground-state
 band at $12_1^+$, corresponding to the backbending.  The relatively
 large $B(E2)$ values between the states with $J$=odd and $J$=even,
 $B(E2:11_1 \rightarrow 12_1)$ and $B(E2:13_1 \rightarrow 12_1)$
 at the $12_1^+$ state are interesting.
 Our model predicts that the odd-spin state $(2J-1)_1^+$ lies near
 above the even-spin state $(2J)_1^+$ when $2J$=10, 12 and 14 but
 the $(2J-1)_1^+$ state is lower than the $(2J)_1^+$ state when
 $2J>14$.  The quality of our model will be examined by the experiment
 on these points.

\begin{table}
\caption{Electric quadrupole properties of the yrast states
 in $^{48}$Cr: ${B(E2)}$ in $e^2$ fm$^4$; $Q_{spec}$ and $Q^{(t)}$
 in $e$ fm$^2$.}
\label{table4}
\begin{center}
\begin{tabular}{rccccccc}\hline
 & Expt.& \multicolumn{3}{c}{present work}
  & \multicolumn{3}{c}{Caurier \textit{et al.}}   \\ 
$J \rightarrow J^\prime$ & $B(E2)$ & $B(E2)$ & Q$_{spec}$
 & Q$^{(t)}_0$ & $B(E2)$ & Q$_{spec}$ & Q$^{(t)}_0$ \\ \hline
$2 \rightarrow 0$ &  321(41)* & 217 & -29.7 & 104
     & 228 & -29.5 & 107 \\
$4 \rightarrow 2$ & 329(110) & 305 & -38.8 & 104
     & 312 & -39.2 & 105 \\ 
$6 \rightarrow 4$ & 301(78)  & 300 & -38.9 &  98
     & 311 & -39.7 & 100 \\ 
$8 \rightarrow 6$ & 230(69)  & 285 & -40.7 &  93
     & 285 & -38.9 &  93 \\ 
$7 \rightarrow 5$ &          &  92 &       &   &   &   &   \\
$10 \rightarrow 8$ & 195(54) & 231 & -35.6 &  83
     & 201 & -22.5 &  77 \\ 
$9 \rightarrow 10$ &         &  38 &       &   &   &   &   \\ 
   $\rightarrow 7$ &         & 124 &       &   &   &   &   \\ 
$12 \rightarrow 10$ & 167(25) & 130 & -11.9 & 62
       & 146 & -5.3 &     \\ 
$11 \rightarrow 12$ &          &  78 &     &   &   &   &   \\ 
   $\rightarrow  9$  &         &  37 &     &   &   &   &   \\ 
   $\rightarrow 10$  &         &   9 &     &   &   &   &   \\ 
$14 \rightarrow 12$& 105(18)   & 105 & -10.8 & 55
       & 116 &      &     \\ 
$13 \rightarrow 11$&           &  43 &     &   &   &   &   \\ 
   $\rightarrow 12$&           &  43 &     &   &   &   &   \\ 
   $\rightarrow 14$&           &  26 & 	   &   &   &   &   \\ 
$15 \rightarrow 13$&           &  49 &     &   &   &   &   \\ 
   $\rightarrow 14$&           &  22 &     &   &   &   &   \\ 
$16 \rightarrow 14$ &  37(8)   &  61 & -12.6 & 42
       &  56 &      &     \\ 
$17 \rightarrow 15$ &          &  69 &     &   &   &   &   \\
   $\rightarrow 16$ &          &  15 &     &   &   &   &   \\ 
$18 \rightarrow 16$ &          & 0.0 & -29.2 & 0.2 &  &  &  \\ 
   $\rightarrow 17$ &          &  16 &     &   &   &   &   \\ \hline
\end{tabular}
\end{center}
\end{table}

 The present model can give information on high-spin states except the
 ground-state band.  Brandolini \textit{et al.} \cite{Brandolini}
 recently identified an energy level $16^+$ at 15.733 MeV.  Near there,
 our model provides two candidates with $16^+$. Cameron \textit{et al.}
 \cite{Cameron3} detected two states $13^+$ and $14^+$ at 10.430 MeV
 and 10.610 MeV before.  One of them possibly corresponds to the
 $13_1^+$ level in the column Calc.  The position of the calculated
 low-lying levels except the yrast states seems to be not so bad.
 The calculated level density is, however, thicker than that observed
 until now.  The present model has a tendency to give too many
 non-yrast levels at low energy, as seen in $^{46}$V.
 This could be attributed not only to the insufficiency of the
 $P_0+P_2+QQ+V^{\tau =0}_{\pi \nu}$ interaction but also to the small
 model space. Since the extension of the model space lowers selectively
 the collective states rather than non-collective states, the full
 $fp$ shell calculation deserves to be tested. 
 
   Figure \ref{fig4} illustrates the isobaric analogue states,
 $4^+$ and $2^+$ with $T=1$ (dotted lines) and $0^+$ with $T=2$
 (dashed line). Our simple interaction including the isoscalar $p$-$n$
 force $V^{\tau=0}_{\pi \nu}$ lays them at roughly good positions,
 if we neglect the inverse order of the $4^+$ and $2^+$ levels in the
 calculated result. 

\subsection{$^{48}$V}

\begin{figure}[b]
\begin{center}
    \epsfig{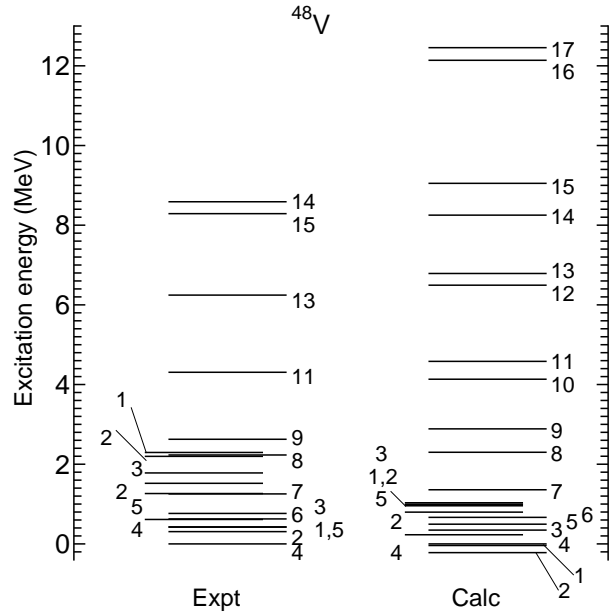}
\caption{Calculated and observed energy levels of $^{48}$V.}
\label{fig5}
\end{center}
\end{figure}

   In Fig. \ref{fig5}, we show calculated energy levels of the odd-odd
 nucleus $^{48}_{23}$V$_{25}$, compared with observed ones
 \cite{ToI,Cameron3}.  Only the yrast levels and 7 other levels with
 definitely assigned spin at low energy are shown in the column Expt,
 and corresponding energy levels obtained by the present model are
 shown in the column Calc.
\begin{table}[b]
\caption{Quadrupole transition probabilities $B(E2)$
 (in $e^2$ fm$^4$) and $Q_{spec}$ (in $e$ fm$^2$) in $^{48}$V.
 The upper part shows calculated $B(E2)$ values of the present work
 and Caurier \textit{et al.}, compared with observed ones. The lower
 part shows the results of present work for high-spin yrast states.}
\label{table5}
\begin{center}
\begin{tabular}{cccccc}\hline
  & Expt. & \multicolumn{2}{c}{present work} & Caurier & \\
$J_n \rightarrow J^\prime_m$ & $B(E2)$ & $B(E2)$ & $Q_{spec}$
  & $B(E2)$ & \\ \hline
     $4_1$             &              &     & 29.5 &       & \\
$2_1 \rightarrow 4_1$  &   28.59(17)  & 105 &      &  48.1 & \\
$5_1 \rightarrow 4_1$  &  104(42)     & 183 & 20.7 & 209.0 & \\
$4_2 \rightarrow 4_1$  &   63(25)     & 182 &      &  28.9 & \\
    $\rightarrow 5_1$  &    $<$41     &  76 &      &  32.0 & \\
$6_1 \rightarrow 5_1$  &  186(73)     & 164 & -0.3 & 191.0 & \\
    $\rightarrow 4_1$  &   46( 6)     &  78 &      &  52.0 & \\
$5_2 \rightarrow 4_2$  &  $<$176(124) &  34 &      &  41.0 & \\
$2_2 \rightarrow 2_1$  &  $>$1.3(19)  & 4.7 &      &  10.7 & \\
       \hline \hline
       \multicolumn{6}{c}{present work} \\
  $J_1$ & $Q_{spec}$ & $\rightarrow J^\prime_1$  & $B(E2)$
   &  $\rightarrow J^\prime_1$ &  $B(E2)$  \\ \hline
$7_1$ &  -6.5 & $\rightarrow 5_1$ &  70 &  $\rightarrow 6_1$ & 114 \\
$8_1$ & -14.0 & $\rightarrow 6_1$ & 138 &  $\rightarrow 7_1$ & 84 \\
$9_1$ & -15.8 & $\rightarrow 7_1$ & 137 &  $\rightarrow 8_1$ & 66 \\
$10_1$ & -21.3 & $\rightarrow 8_1$ & 137 & $\rightarrow 9_1$ & 51 \\
$11_1$ & -15.6 & $\rightarrow 9_1$ & 128 & $\rightarrow 10_1$ & 31 \\
$12_1$ & -23.7 & $\rightarrow 10_1$ & 95 & $\rightarrow 11_1$ & 22 \\
$13_1$ & -17.7 & $\rightarrow 11_1$ & 95 & $\rightarrow 12_1$ & 18 \\
$14_1$ &  -7.9 & $\rightarrow 12_1$ & 24 & $\rightarrow 13_1$ & 2.2 \\
$15_1$ & -13.7 & $\rightarrow 13_1$ & 48 & $\rightarrow 14_1$ & 11  \\
$16_1$ & -18.2 & $\rightarrow 14_1$ & 2.3& $\rightarrow 15_1$ & 9.1 \\
$17_1$ & -22.8 & $\rightarrow 15_1$ & 1.5& $\rightarrow 16_1$ & 5.7 \\
 \hline
\end{tabular}
\end{center}
\end{table}
  The present model gives lower energies to the $2_1^+$ and $1_1^+$
 states than the $4_1^+$ ground state.  The states except the yrast
 states come to low energy region as compared with the observed ones.
  We have already stated the overbinding of non-collective states.  
 The $P_0 + P_2 + QQ + V_{\pi \nu}^{\tau=0}$ interaction is still able
 to describe the yrast band $4_1^+$, $5_1^+$, $6_1^+$ ... $15_1^+$
 based on the $4_1^+$ state as seen in Fig. \ref{fig5}. 
 The correspondence between calculated and observed yrast levels is 
 quite well.   The quality is the same as that of the full $fp$ shell
 model calculation with the KB3 interaction \cite{Caurier}.  
 There is, however, a discrepancy between theory and experiment about
 the order of the levels $14_1^+$ and $15_1^+$.

   Table \ref{table5} shows theoretical and experimental values of
 $B(E2)$ in $^{48}$V.  The present results on the $B(E2)$ values
 agree well with those of the full $fp$ shell model calculation
 with the KB3 interaction \cite{Caurier}, except the $2_1^+$ and
 $4_2^+$ states. Although there are little experimental data, our
 model seems to be able to explain the observed large values of
 $B(E2)$ between the yrast states $4_1^+$, $5_1^+$, $6_1^+$ etc.
 to the same extent as the full $fp$ shell model calculation of
 Caurier \textit{et al.}
 The calculated results suggest a structure change in the yrast band
 at $8_1^+$.  The spectroscopic quadrupole moment $Q_{spec}$ changes
 rapidly from the positive value at the $4_1^+$ state to the negative
 value at the $8_1^+$ state.  Coincidently, $B(E2:J \rightarrow J-1)$
 is larger than $B(E2:J \rightarrow J-2)$ up to $7_1^+$, while
 $B(E2:J \rightarrow J-2)$ is larger than $B(E2:J \rightarrow J-1)$
 from $8_1^+$ till $15_1^+$. There seem to be two series of states with
 even $J$ and odd $J$ connected by the large $B(E2)$ above the $7^+_1$
 level, which is consistent with the observed strong $E2$ transitions
 $15_1^+ \rightarrow 13_1^+ \rightarrow 11_1^+ \rightarrow 9_1^+$ (the
 calculated $Q_{spec}$ values is nearly constant for these states).
  This is similar to the case of $^{46}$V.  There is, however,
 a difference between $^{46}$V and $^{48}$V in the calculated results.
 The order of the two levels $(2J)_1^+$ and $(2J+1)_1^+$ is reversed
 for $2J \geq 8$ in $^{46}$V, while it is in order of spin in $^{48}$V
 except $14_1^+$ and $15_1^+$.  Future experiments will judge the
 quality of our model.  The discrepancy between theory and experiment
 \cite{Cameron3} with respect to the order of the two levels $14_1^+$
 and $15_1^+$ should be examined further.

\subsection{$^{50}$Cr}

\begin{figure}[b]
\begin{center}
    \epsfig{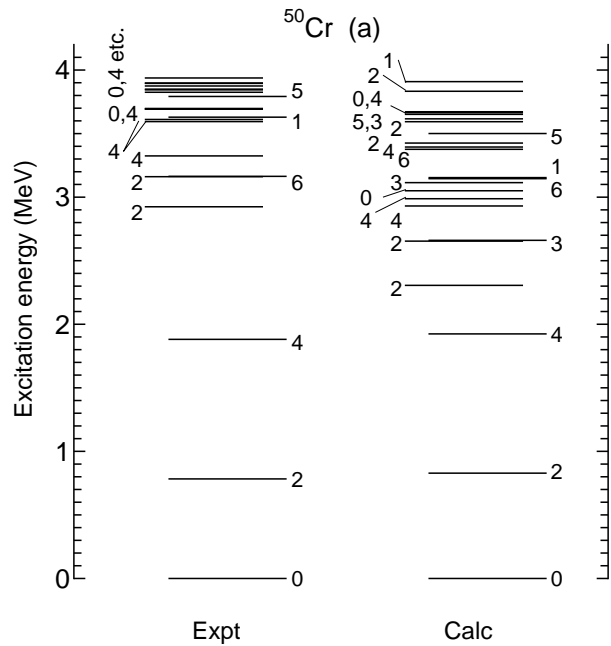}
\caption{Calculated and observed energy levels of $^{50}$Cr.
         The states below 4 MeV are shown.}
\label{fig6}
\end{center}
\end{figure}

\begin{figure}[t]
\begin{center}
    \epsfig{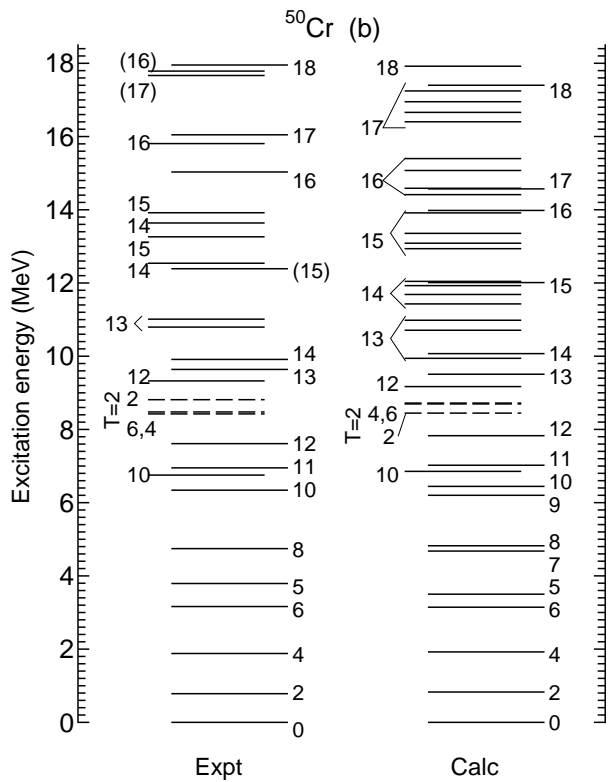}
\caption{Calculated and observed yrast levels, high-spin states and
         $T=2$ isobaric analogue states of $^{50}$Cr.}
\label{fig7}
\end{center}
\end{figure}

   Let us compare calculated energy levels of $^{50}$Cr with 
 observed ones \cite{ToI,Cameron2,Brandolini,Lenzi} in detail.
 Figure \ref{fig6} shows low-lying states below 4 MeV and Fig.
 \ref{fig7} illustrates a gross level scheme, mainly the yrast states
 and high-spin states.  In Table \ref{table6}, we compare electric 
 quadrupole properties between theory and experiment.  The observed
 values of $B(E2)$ are from Refs. \cite{Brandolini} (Expt.1) and
 \cite{Cameron2} (Expt.2) and $B(E2:2_1^+  \rightarrow 0_1^+)$ is
 from Ref. \cite{Burrows2}. The theoretical results of
 Mart\'{\i}nez-Pinedo \textit{et al.} \cite{Martinez} and Zamick
 \textit{et al.} \cite{Zamick} are listed for comparison.

   Figures \ref{fig6}, \ref{fig7} and Table \ref{table6} demonstrate
 that the present model excellently describes the energies and electric
 quadrupole properties of the low-spin collective states $0_1^+$,
 $2_1^+$, $4_1^+$, $6_1^+$ and $8_1^+$ before the backbending in the
 ground-state band. (Remember that the ground-state energy is
 reproduced very well.)  Figure \ref{fig6}, at the same time, reveals
 the insufficiency of the present model for non-collective states.
 The calculated low-lying states except the collective states lie
 lower than the observed levels, though there seems to be a certain
 correspondence between theory and experiment.

   As confirmed in Fig. \ref{fig7}, the correspondence between theory
 and experiment is very well for the ground-state band, except for the
 $16^+$ level.  The calculated $E2$ transition probabilities
 $B(E2:J \rightarrow J-2)$ in the ground-state band, which are
 comparable to those of the full $fp$ shell model calculation with the
 KB3 interaction, well explain the observed $B(E2)$ values except that
 our model suggests a structure change at $10^+_1$ by the small value
 of $B(E2:10_1^+ \rightarrow 8_1^+)$.  The spectroscopic quadrupole
 moment $Q_{spec}$, which is also comparable to that of
 Mart\'{\i}nez-Pinedo \textit{et al.}, abruptly changes to the positive
 value at $10_1^+$ from the negative values up to $8_1^+$, indicating
 the structure change.
 Our model, on the other hand, reproduces the large value of
 $B(E2:10_2^+ \rightarrow 8_1^+)$ as well as the energy of the
 $10_2^+$ state.  The nice reproduction of the energy levels $8_1^+$,
 $10_1^+$, $10_2^+$, $11_1^+$, $12_1^+$, $12_2^+$, $13_1^+$ and
 $14_1^+$ manifests that our model describes the backbending phenomenon
 of the ground-state band well. It is interesting that the $E2$
 transition probability $B(E2:J \rightarrow J-1)$ is larger than
 $B(E2:J \rightarrow J-2)$ for the $11_1^+$, $12_1^+$ and $13_1^+$
 states near the backbending.
      
\begin{table}
\caption{Electric quadrupole properties of the yrast and other states
 in $^{50}$Cr: ${B(E2)}$ in $e^2$ fm$^4$; $Q_{spec}$, $Q^{(s)}$ and
 $Q^{(t)}$ in $e$ fm$^2$. }
\label{table6}
\begin{center}
\begin{tabular}{rcccccccccc}\hline
  & \multicolumn{2}{c}{$B(E2)$} & \multicolumn{4}{c}{present work}
   & \multicolumn{2}{c}{Mart\'{\i}nez \textit{et al.}}
   &  \multicolumn{2}{c}{Zamic \textit{et al.}}           \\
 $J_n \rightarrow J^\prime_m$ & Expt.1 & Expt.2
  & $B(E2)$ & $Q_{spec}$ & $Q_0^{(s)}$ & $Q_0^{(t)}$
  & $B(E2)$ & $Q_{spec}$ & $B(E2)$ & $Q_{spec}$  \\ \hline
 $2_1 \rightarrow 0_1$ & \multicolumn{2}{c}{217$\pm$25}& 215 & -29
  & 101 & 104   & 193 & -27   & 173 & -24.8 \\
 $4_1 \rightarrow 2_1$ & 204$\pm$57 &           & 302 & -38
  & 103 & 103   & 264 & -33   & 246 & -30.0 \\
 $6_1 \rightarrow 4_1$ & 235$\pm$47 &           & 211 & -18
  &  44 &  82   & 224 & -18   & 215 & -15.6 \\
 $8_1 \rightarrow 6_1$ & 205$\pm$51 &           & 200 & -25
  &  59 &  78   & 200 & -17   & 192 & -14.7 \\
 $8_2 \rightarrow 6_1$ &            &           & 0.3 & 4.3
  &     &       &     &       & 0.4 & 19.5  \\
 $10_1 \rightarrow 8_1$ & 72$\pm$14 & 66$^{+102}_{-34}$ & 14 & 47
  & -108 & 20   &  54 &  30   &  81 & 26.5  \\
      $\rightarrow 8_2$ &           &           &  15 &    
        &    &    &    &      &  32 &       \\
      $\rightarrow 9_1$ &           &           & 2.7 &    
        &    &    &    &      &     &       \\
 $10_2 \rightarrow 8_1$ & 131$\pm$26 &          & 145 & -14
        &    &    &    &      &  68 & 12.5  \\
      $\rightarrow 8_2$ &           &           & 0.7 &    
        &    &    &    &      & 9.5 &       \\
 $11_1 \rightarrow 9_1$ &           &           &  41 & 29
        &    &    &    &      &     &       \\
      $\rightarrow 10_1$ &          &           &  78 &    
        &    &    &    &      &     &       \\
 $12_1 \rightarrow 10_1$ & 52$\pm$7 &    0+23   &  29 & 17
        & -37 & 29 &  30 & 13  &  69 & 13.0  \\
      $\rightarrow 10_2$ &          &           &   9 &   
        &    &    &      &     &  12 &       \\
      $\rightarrow 11_1$ &          &           &  51 &   
        &    &    &      &     &     &       \\
 $12_2 \rightarrow 10_1$ &          &           & 1.0 & -17
        &    &    &      &     & 0.0 & -6.6  \\
      $\rightarrow 10_2$ &          &           &  20 &    
        &    &    &      &     &  51 &       \\
 $13_1 \rightarrow 11_1$ &   $>$5   &   $>$30   &  42 & 8.6
        &    &    &      &     &     &       \\
      $\rightarrow 12_1$ &          &           &  49 &    
        &    &    &      &     &     &       \\
 $14_1 \rightarrow 12_1$ &  43$\pm$7 &  31$\pm$7 &  52 & 10
        & -22 & 39  & 60 &  8  &  67 &  8.2  \\
      $\rightarrow 13_1$ &          &            &  13 &   
        &     &     &    &     &     &       \\
 $15_1 \rightarrow 13_1$ &  $>$10   &    $>$21   &  11 & -6.4
        &    &      &    &     &     &       \\
      $\rightarrow 14_1$ &          &            & 0.8 &   
        &    &      &    &     &     &       \\
 $16_1 \rightarrow 14_1$ &  $>$10   &            & 7.8 & 7.6
        & -17 & 15  &  3 & 10  &  .8 & -8.7  \\
      $\rightarrow 15_1$ &          &            & 3.3 &    
        &    &      &    &     &     &       \\
 $17_1 \rightarrow 15_1$ &          &    $>$66   & 67 & -6.2
        &    &      &    &     &     &       \\
      $\rightarrow 16_1$ &          &            & 14 &    
        &    &      &    &     &     &       \\
 $18_1 \rightarrow 16_1$ &  50$\pm$20 &          & 0.5 & -25
        & 53 & 3.9  & 76 &  9  &  83 & -8.0  \\
      $\rightarrow 17_1$ &          &            & 17 &   
        &    &      &    &     &      &      \\ \hline
\end{tabular}
\end{center}
\end{table}

  There is a disagreement between theory and experiment on the 
 energies of the $16_1^+$ and $17_1^+$ states and 
 $B(E2:18_1^+ \rightarrow 16_1^+)$.
 A little possibility that the two levels $16^+$ and $17^+$ observed
 at 15.032 MeV and 16.048 MeV are not the yrast states remains.
 Our model predicts a bundle of levels with the same $J$ near each 
 high-spin yrast level.  Actually, there are calculated energy levels 
 corresponding well to the observed levels with $10^+$, $12^+$,
 $13^+$, $14^+$ and $15^+$ in addition to the yrast levels.  
 The observed $16^+$ and $17^+$ states at 15.032 MeV and 16.048 MeV
 have such candidates at the same energy region in the calculated
 energy spectra.
 The calculated values $B(E2:18_1^+ \rightarrow 16_1^+)=0.5$ and
 $B(E2:18_2^+ \rightarrow 16_1^+)=1.7$ in $e^2$ fm$^4$ are much smaller
 than the observed one. The absence of the $f_{5/2}$ orbit may have
 the influence on the energies and $B(E2)$ of the high-spin states
 $16^+$, $17^+$ and $18^+$, in the present model.

   Our model reproduces the isobaric analogue states $6^+$, $4^+$ and
 $2^+$ with $T$=2 (broken lines) at good energy, though the order of
 the calculated levels $(6^+, 4^+)$ and $2^+$ is reverse to that of
 the observed ones. This is a nice work of the isoscalar $p$-$n$ force
 $V_{\pi \nu}^{\tau=0}$.

\subsection{$^{50}$Mn}

\begin{figure}[b]
\begin{center}
    \epsfig{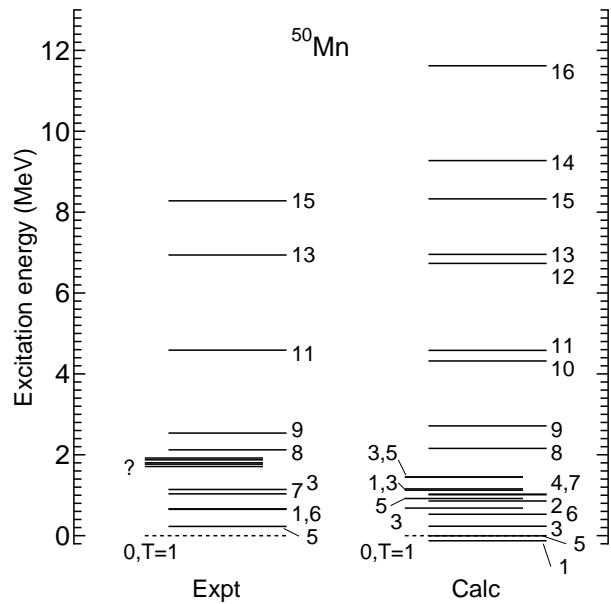}
\caption{Calculated and observed energy levels of $^{50}$Mn.}
\label{fig8}
\end{center}
\end{figure}

\begin{table}[b]
\caption{Electric quadrupole properties of the yrast states in
 $^{50}$Mn: $B(E2)$ in $e^2$ fm$^4$ and various $Q$ values in
 $e^2$ fm$^2$.}
\label{table7}
\begin{center}
\begin{tabular}{rcccccccc} \hline
 &  \multicolumn{4}{c}{present work}
 &  \multicolumn{4}{c}{Svenssen \textit{et al.}} \\
 $J \rightarrow J^\prime$ & $B(E2)$ & $Q_{spec}$ & $Q^{(s)}_0$
  & $Q^{(t)}_0$ & $B(E2)$ & $Q_{spec}$ & $Q^{(s)}_0$ & $Q^{(t)}_0$
   \\ \hline   
 3 $\rightarrow$ 1  & 259 & -33 & -19 &   & 240 &     &    &    \\
 5 \ \ \            &     & 59 & 102 &    &     & 57  & 98 &    \\
 6 $\rightarrow$  5 & 294 & 32 & 103 & 99 & 258 & 32 & 102 & 93 \\
 7 $\rightarrow$  6 & 301 & 13 &  94 & 92 & 251 & 14 & 100 & 84 \\
   $\rightarrow$  5 &  39 &    &     & 90 &  42 &    &     & 92 \\
 8 $\rightarrow$  7 & 276 & 3.4 & 191 & 90 & 140 & 5 & 285 & 64 \\
   $\rightarrow$  6 &  89 &    &     & 95 &  74 &    &     & 87 \\
 9 $\rightarrow$  8 & 217 & -5.6 & 78 & 84 & 142 & -0.6 & 9 & 68 \\
   $\rightarrow$  7 & 127 &    &     & 95 & 133 &    &     & 97 \\
10 $\rightarrow$  9 & 177 & -8.1 & 59 & 81 &     &   &     &    \\
   $\rightarrow$  8 & 128 &    &     & 85 &     &    &     &    \\
11 $\rightarrow$  9 &  97 & 5.4 & -28 & 69 & 130 & 0.5 & -2.6 & 80 \\
   $\rightarrow$ 10 &  62 &    &     & 51 &     &    &     &    \\
12 $\rightarrow$ 11 & 119 & -0.4 & 1.5 & 76 &   &    &     &    \\
   $\rightarrow$ 10 &  73 &    &     & 57 &    &     &     &    \\
13 $\rightarrow$ 11 &  83 & -1.5 & 5.6 & 58 & 99 & 6 & -23 & 64 \\
   $\rightarrow$ 12 &  61 &    &     & 59 &    &     &     &    \\
14 $\rightarrow$ 13 &  69 & -1.8 & 6.2 & 65 &    &    &     &    \\
   $\rightarrow$ 12 &  43 &    &     & 41 &    &     &     &    \\
   $\rightarrow$ 15 & 0.0 &    &     &    &    &     &     &    \\
15 $\rightarrow$ 13 &  27 & 33 & -105 & 32 & 47 & 25 & -80 & 41 \\
16 $\rightarrow$ 15 &  43 & 14 & -42 & 58 &    &    &     &    \\
   $\rightarrow$ 14 &  47 &    &     & 41 &    &     &     &  \\ \hline
\end{tabular}
\end{center}
\end{table}

   A recent experiment by Svensson \textit{et al.} \cite{Svensson}
 identified high-spin states in the $N=Z$ odd-odd nucleus
 $^{50}_{25}$Mn$_{25}$. We compare the energy scheme obtained by the
 present model with the observed one in Fig. \ref{fig8}. Like $^{48}$V,
 our model lays another $T$=0 state $(1_1^+)$ at lower energy than the
 $5_1^+$ state which is the lowest $T$=0 state in experiment.
  Moreover, these two calculated states $1^+_1$ and $5^+_1$ lie lower
 than the ground state $0_1^+$ with $T$=1.  The discrepancy between
 theory and experiment for the $T$=0, $5_1^+$ level, however, is only
 0.23 MeV. For such a simple model, the obtained energies of the $T$=1,
 $0^+$ and $T$=0, $5_1^+$ states are rather satisfactory.
 We see again that the calculated non-collective states except the
 yrast states lie lower than the observed ones.

   It is impressive that the observed $T$=0 yrast band on the $5_1^+$
 state is reproduced so well by our simple model.  The correspondence
 between theory and experiment is better than that of the full $fp$
 shell model calculation using the KB3 interaction and single-particle
 energies taken from $^{41}$Ca \cite{Svensson}. Calculated $E2$
 transition probabilities between the yrast states are compared with
 those of Ref.\cite{Svensson} in Table \ref{table7}. The $B(E2)$
 values in the present work are larger up to $9_1^+$ and smaller above
 $10_1^+$ than those of Ref. \cite{Svensson}.

   The theoretical $B(E2)$ values between the yrast states in 
 $^{50}$Mn show a different feature from those in $^{46}$V and
 $^{48}$V.  Namely, every $B(E2:J \rightarrow J-1)$ is larger than
 $B(E2:J \rightarrow J-2)$ except the $15_1^+$ state in $^{50}$Mn,
 while the relative magnitudes of $B(E2:J \rightarrow J-1)$ and
 $B(E2:J \rightarrow J-2)$ are reversed above $7_1^+$ corresponding
 to the structure change in $^{46}$V and $^{48}$V.  In $^{50}$Mn, the
 spectroscopic quadrupole moment $Q_{spec}$ changes one after another
 but the intrinsic quadrupole moment $Q_0^{(t)}$ keeps relatively
 large values up to high spin. There is probably a delicate difference
 between $^{46}$V and $^{50}$Mn which are cross-conjugate systems
 $(f_{7/2})^{6p}$ and $(f_{7/2})^{6h}$ within the single $j$ model
 space.  The present result suggests a considerably large contribution
 of the upper orbits $p_{3/2}$ and $p_{1/2}$ in $^{50}$Mn.  Our model
 yields different level schemes in $^{46}$V and $^{50}$Mn. The inverse
 order of the levels $(2J)_1^+$ and $(2J+1)_1^+$ happens above $2J=8$
 in $^{46}$V but happens only at $2J=14$ in $^{50}$Mn.
 This interesting difference will be examined by detecting the
 even-$J$ high-spin states.

\section{Discussions}

\subsection{Dependence of the calculated results on the model}

\begin{figure}
\begin{center}
    \epsfig{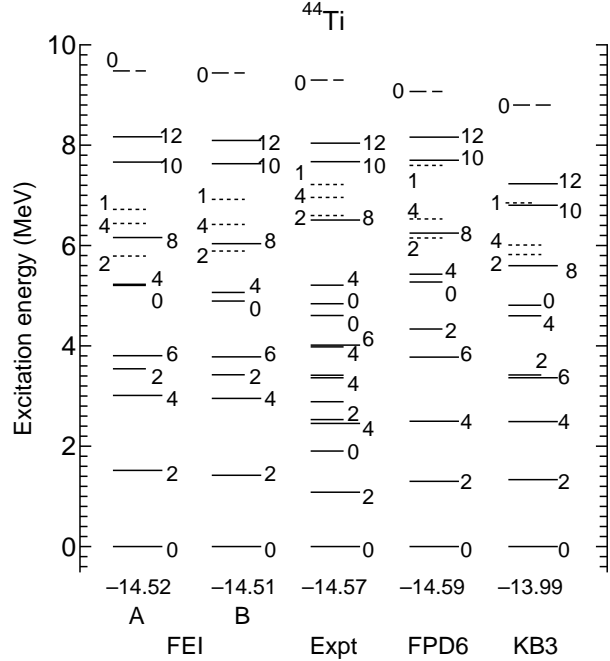}
\caption{Energy levels obtained using various effective interactions
         in $^{44}$Ti, compared with observed levels.  The ground-state
         energy is written below every $0^+_1$ level in MeV.
         The dotted lines denote the $T=1$ states and dashed lines
         denote the $T=2$, $0^+$ state. The space A is
         ($f_{7/2},p_{3/2},p_{1/2}$) and B is the full $fp$ space.}
\label{fig9}
\end{center}
\end{figure}

@We have seen the success of the present model in the small model
 space $(f_{7/2}, p_{3/2}, p_{1/2})$.  To see the dependence on the
 model space, we made calculations with our interaction in the full
 $fp$ shell for $^{44}$Ti where the calculations are easy.  We took
 $\epsilon(f_{5/2})$=4.88 MeV from $^{41}$Ca and changed the force 
 strengths as little as possible so as to reproduce the observed 
 energy levels of $^{44}$Ti,
  i.e., $g_0=0.42(42/A)$, $\chi^\prime=0.29(42/A)^{5/3}$, and $g_2$
 and $k^0$ being unchanged.  Obtained energy levels (column B) are
 compared with those in the $(f_{7/2}, p_{3/2}, p_{1/2})$ space
 (column A) and also with the results obtained using the FPD6
 \cite{Richter} and KB3 interactions in the full $fp$ space, in Fig.
 \ref{fig9}. This figure teaches that the extension of the model
 space to the full $fp$ space modifies little the energy levels of the
 collective states in our model. This is probably true for the
 collective states of $A$=46, 48 and 50 nuclei, as guessed from the
 successful results in Sec. III.  We shall see the reason in the next
 subsection.

  According to Fig. \ref{fig9}, our interaction FEI is roughly
 comparable to the FPD6 and KB3 interactions also in $^{44}$Ti.
 The KB3 interaction yields rather compressed energy levels for the
 ground-state band and a shallow binding energy.  This could be
 attributed to the lack of the $g_{9/2}$ orbit which is included in
 the model space of the original Kuo-Brown (KB) interaction \cite{Kuo}.
 If the $g_{9/2}$ orbit is added, the additional interactions such as
 the pairing force will lower the low-lying collective states $0_1^+$,
 $2_1^+$ ... and recover the binding energy.  This situation presents
 a contrast to the success of the KB3 interaction in the
 $A$=46, 48 and 50 nuclei.
     
\begin{table}[t]
\caption{Dependence of $B(E2:J^+_1 \rightarrow (J-2)^+_1)$
 on model spaces and effective interactions in $^{44}$Ti.
 The space A is ($f_{7/2}, p_{3/2}, p_{1/2}$) and B is the full
 $fp$ space. The $B(E2)$ values obtained by the $\alpha$-cluster
 model is also shown.}
\label{table8}
\begin{center}
\begin{tabular}{cccccccccc}\hline
      & \multicolumn{2}{c}{FEI} & Expt. & \multicolumn{2}{c}{FPD6}
        & \multicolumn{2}{c}{KB3} & $\alpha$-        \\
  $J$ & A   &   B &     & A & B & A & B & cluster        \\ \hline
    2 & 102 & 103 & 120$\pm$37 & 80 & 60 & 73 & 64 & 107 \\
    4 & 125 & 134 & 277$\pm$55 &101 & 72 & 94 & 75 & 146 \\
    6 &  82 &  74 & 157$\pm$28 & 99 & 63 & 91 & 57 & 140 \\
    8 &  77 &  73 &     $>18$  & 91 & 58 & 84 & 50 & 118 \\
   10 &  76 &  67 & 138$\pm$28 & 79 & 67 & 70 & 62 &  75 \\
   12 &  52 &  52 &  40$\pm$8  & 50 & 52 & 44 & 47 &  34 \\ \hline
\end{tabular}
\end{center}
\end{table}

  The extension of the model space, of course, changes the wave
 functions.  Table \ref{table8} lists the
 $B(E2:J_1^+ \rightarrow (J-2)_1^+)$ values calculated in the
 $(f_{7/2}, p_{3/2}, p_{1/2})$ space and full $fp$ space, which are
 compared with the observed $B(E2)$ values and those obtained by the
 $\alpha$-cluster model \cite{Michel}.  The extension of the model
 space enlarges the $B(E2)$ values between the most collective states
 $2_1^+$ and $4_1^+$, when our interaction is used.  The same does
 not happen for the realistic effective interactions FPD6 and KB3.
 The two interactions rather reduce the $B(E2)$ values between the
 $2_1^+$, $4_1^+$, $6_1^+$ and $8_1^+$ states.  An advantage of the
 $\alpha$-cluster  model \cite{Michel} was the reproduction of the
 large $B(E2)$ values without the effective charge as compared with
 the reduced $B(E2)$ values of the shell models using the realistic
 effective interactions.  The shell model using the FEI yields the
 $B(E2)$ values comparable with the $\alpha$-cluster model,
 though the former employs the effective charge $e_{eff}=0.5e$.
 The $^{44}$Ti nucleus is more or less in an $\alpha$-like four-nucleon
 correlated state outside the doubly-closed-shell $^{40}$Ca core
 \cite{Ohkubo}.  The large $B(E2)$ values are related to the
 $\alpha$-like four-nucleon correlations.
 The $P_0 + P_2 + QQ + V_{\pi \nu}^{\tau=0}$ interaction, which is
 suitable for the description of the nuclear collective motions
 (the quadrupole vibration and deformation), is considered to have
 an affinity also for the $\alpha$-like four-nucleon correlations
 \cite{Hasegawa2} as compared with the FPD6 and KB3 interactions
 as seen in Table \ref{table8}.
 Anyhow, our interaction including the $QQ$ force can give larger
 $B(E2)$ values than the FPD6 and KB3 interactions, when the same
 model space is used.  In the configuration
 $(f_{7/2}, p_{3/2}, p_{1/2})^8$ for $^{48}$Cr, for instance, the KB3
 interaction gives $B(E2:4^+_1 \rightarrow 2^+_1)=196$ $e^2$ fm$^4$,
 while the FEI gives $B(E2:4^+_1 \rightarrow 2^+_1)=305$ $e^2$ fm$^4$.
 Thus our interaction in the smaller model space
 $(f_{7/2}, p_{3/2}, p_{1/2})$ reproduces large $B(E2)$ values
 comparable to the FPD6 and KB3 interactions in the full $fp$ space,
 in the $A$=46, 48 and 50 nuclei.  If we extend the model space
 using our interaction, a smaller effective charge will serve in the
 $A$=46-50 nuclei.

\begin{table}[b]
\caption{Effects of varying the single-particle energies on the
 excitation energies $E_x$ (or ground-state energies $E_{gs}$)
 and $B(E2)$ values.  The single-particle energies
 $(\epsilon_{7/2}, \epsilon_{3/2}, \epsilon_{1/2})$ are taken from
 $^{41}$Ca in I and are Kuo-Brown's in II, respectively.}
\label{table9}
\begin{center}
\begin{tabular}{rcccc}\hline
 &  \multicolumn{2}{c}{$E_x$ or ($E_{gs}$)} 
 & \multicolumn{2}{c}{$B(E2:J \rightarrow J-2)$} \\
 $A$   $J^\pi_1$  &    I      &    II     &   I   &   II  \\ \hline
$^{46}$Ti $0^+_1$ & (-20.169) & (-19.987) &       &       \\
          $2^+_1$ &   1.110   &  1.133    & 119 & 116 \\
 	  $4^+_1$ &   2.212   &  2.220    & 159 & 152 \\
  	  $6^+_1$ &   3.237   &  3.198    & 141 & 133 \\
	  $8^+_1$ &   4.984   &  4.934    & 126 & 119 \\
	  $10^+_1$ &  6.570   &  6.479    & 104 & 100 \\
	  $11^+_1$ &  7.548   &  7.420    &  51 &  48 \\
	  $12^+_1$ &  8.177   &  8.049    &  53 &  51 \\
	  $14^+_1$ & 10.491   & 10.330    &  26 &  25 \\ \hline
$^{48}$Cr $0^+_1$ & (-32.380) & (-32.131) &       &       \\
	  $2^+_1$ &   0.785   &  0.802    & 217 & 211 \\
	  $4^+_1$ &   1.837   &  1.847    & 306 & 297 \\
	  $6^+_1$ &   3.222   &  3.217    & 301 & 287 \\
	  $7^+_1$ &   5.554   &  5.524    &  92 &  89 \\
	  $8^+_1$ &   4.887   &  4.866    & 285 & 274 \\
	  $10^+_1$ &  6.670   &  6.606    & 232 & 219 \\
	  $12^+_1$ &  8.238   &  8.095    & 130 & 127 \\
	  $13^+_1$ & 10.878   & 10.770    &  43 &  32 \\
	  $14^+_1$ & 10.429   & 10.248    & 105 & 103 \\ \hline
\end{tabular}
\end{center}
\end{table}

  There is another sign that the three interactions FEI, FPD6 and KB3
 give different wave functions.  For $^{44}$Ti, the expectation
 values of nucleon number in respective orbits of the $fp$ space are
 as follows: $n_{7/2}$=3.082, $n_{3/2}$=0.701, $n_{1/2}$=0.103,
 $n_{5/2}$=0.114 for FEI; $n_{7/2}$=3.061, $n_{3/2}$=0.464,
 $n_{1/2}$=0.124, $n_{5/2}$=0.351 for FPD6; $n_{7/2}$=3.367,
 $n_{3/2}$=0.262, $n_{1/2}$=0.085, $n_{5/2}$=0.286 for KB3.
 The contribution of the $p_{3/2}$ orbit is large in FEI.  
 Nucleons are distributed over all the orbits most in FPD6.
 It should be noticed that interactions related to the $f_{7/2}$ orbit
 are strengthened in KB3 by the modification from the KB interaction
 \cite{Caurier}, resulting in the large value of $n_{7/2}$.

   The results of calculations depend also on the single-particle 
 energies.  We examined the dependence by shell model calculations
 using Kuo-Brown's single-particle energies, $\epsilon_{7/2}$=0.0,
 $\epsilon_{3/2}$=2.1 and $\epsilon_{1/2}$=3.9 in MeV.
 In Table \ref{table9}, calculated energies and $B(E2)$ for the
 ground-state bands of $^{46}$Ti and $^{48}$Cr are compared with
 those of Sec. III (where $\epsilon_{7/2}$=0.0, $\epsilon_{3/2}$=1.94
 and $\epsilon_{1/2}$=3.61 in MeV). Table \ref{table9} says that
 the change of the single-particle energies tried here does
 not significantly affect the energies and $B(E2)$.  We, therefore,
 need not change the basic understanding in Sec. III, as long as we
 use ordinary single-particle energies.

\subsection{Characteristics of the interaction matrix elements}

\begin{figure}[b]
\begin{center}
    \epsfig{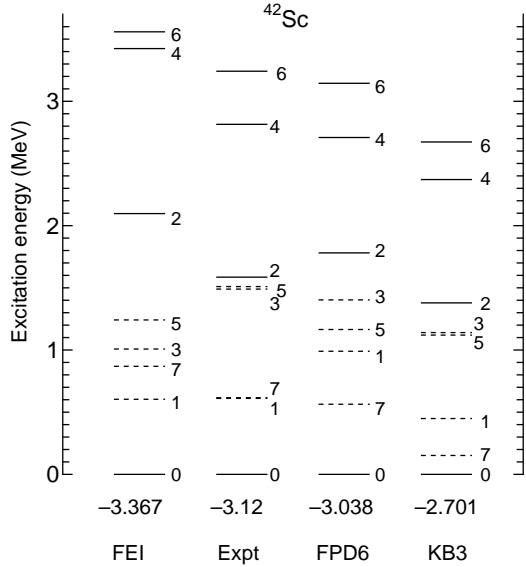}
\caption{Energy levels obtained using various effective interactions
         in $^{42}$Sc, compared with observed levels.  The ground-state
         energy is written below every $0^+_1$ level in MeV.
         The solid lines denote the $T=1$ states and dotted lines
         denote the $T=0$ states.}
\label{fig10}
\end{center}
\end{figure}

   The effective interactions can be directly checked in $^{42}$Sc.
 In Fig. \ref{fig10}, we compare energy levels obtained using the
 three interactions (FEI, FPD6 and KB3) in $^{42}$Sc with observed
 energy levels, where the model space $(f_{7/2}, p_{3/2}, p_{1/2})$ 
 is employed for FEI.  This figure shows that the isoscalar ($T=0$) 
 $p$-$n$ interactions of FEI are best but the isovector $(T=1)$ 
 interactions of FEI are worst among the three interactions. 
  We now understand the reason why our interaction is not good for
 the Ca isotopes where only the isovector interactions take action.
 Probably, the isovector interactions play dominant roles in low-lying
 states of nuclei with $n_p$=1 or $n_n$=1. The success of our
 interaction FEI for the collective states in the $A$=46, 48 and 50
 nuclei with $n_p \ge 2$ and $n_n \ge 2$ suggests leading roles
 of the isoscalar $p$-$n$ interactions there. 

   The compression of the $T=1$ energy levels $(0^+, 2^+, 4^+,6^+)$
 in the KB3 result of Fig. \ref{fig10} is due to the lack of the
 $g_{9/2}$ orbit (compare with the figure of Ref. \cite{Kuo}), as
 mentioned in the previous subsection.  The KB3 interaction lays
 the $T=0$ states $(7^+, 1^+, 5^+, 3^+)$ lower than the FPD6.
 This is caused by the modification of interaction matrix elements
 from the original KB interaction \cite{Kuo} to the KB3
 \cite{Caurier}, where the most important modification is to
 strengthen the isoscalar $p$-$n$ interactions.

  In Table \ref{table10}, we list typical interaction matrix elements
 of the four interactions FPD6, KB, KB3 and FEI.  For FEI, the force
 parameters fixed in $^{44}$Ti are applied to the $A=42$ system, i.e.,
 $g_0=0.42$, $g_2=0.36$, $\chi^\prime=0.30$ and $k^0=2.23$ in MeV.
 Let us denote diagonal matrix elements as
 $V(abJT)=\langle abJT|V|abJT \rangle$.
 The KB3 interaction enlarges $V(f_{7/2} f_{7/2} J=odd,T=0)$
 more than $V(f_{7/2} f_{7/2} J=even,T=1)$, especially
 $V(f_{7/2} f_{7/2} J=1, T=0)$ and $V(f_{7/2} f_{7/2} J=3, T=0)$
 as well as $V(f_{7/2} r J, T=0)$ with
 $r=(p_{3/2}, p_{1/2}, f_{5/2})$. 
 These modifications bring the KB3 interaction a little near to the
 FEI.  In our interaction FEI, the $V_{\pi \nu}^{\tau=0}$ force
 strengthens just the isoscalar ($T=0$) $p$-$n$ interactions, and
 the matrix elements $V(f_{7/2} f_{7/2} J=1, T=0)$ and
 $V(f_{7/2} f_{7/2} J=3, T=0)$ are comparable with
 $V(f_{7/2} f_{7/2} J=7, T=0)$, which makes the $1^+$ state lowest
 among the $T=0$ states of $^{42}$Sc.
 
\begin{table}
\caption{Comparison of interaction matrix elements
 $\langle abJT|V|cdJT \rangle$ between various effective interactions.
 The orbits are labeled by numbers, 1: $f_{7/2}$, 2: $p_{3/2}$,
 3: $p_{1/2}$, 4: $f_{5/2}$.}
\label{table10}
\begin{center}
\begin{tabular}{cccccc}\hline
$a \ \ b \ \ c \ \ d$ & $J$, $T$ & FPD6 & KB   & KB3    & FEI
    \\ \hline
1 1 1 1   & 1, 0 & -0.177 & -0.525 & -1.175 & -2.680 \\
          & 3, 0 & -0.499 & -0.208 & -0.858 & -2.257 \\
          & 5, 0 & -1.046 & -0.502 & -0.852 & -1.939 \\
          & 7, 0 & -2.474 & -2.199 & -2.549 & -2.490 \\ \hline
1 1 1 1   & 0, 1 & -2.268 & -1.807 & -1.917 & -2.236 \\
          & 2, 1 & -0.888 & -0.785 & -1.095 & -0.812 \\
          & 4, 1 & -0.144 & -0.087 & -0.197 &  0.185 \\
          & 6, 1 &  0.168 &  0.226 &  0.116 &  0.185 \\ \hline
1 2 1 2   & 2, 0 &  0.003 & -0.298 & -0.593 & -2.552 \\
          & 3, 0 & -0.823 & -0.604 & -0.904 & -2.178 \\
          & 4, 0 & -0.547 & -0.164 & -0.464 & -1.763 \\
          & 5, 0 & -2.522 & -2.165 & -2.465 & -2.801 \\ \hline
1 3 1 3   & 3, 0 & -1.824 & -1.484 & -1.784 & -2.230 \\
          & 4, 0 & -0.787 & -0.746 & -1.046 & -2.230 \\ \hline
1 4 1 4   & 1, 0 & -4.665 & -3.621 & -3.921 & -2.716 \\
          & 2, 0 & -2.950 & -2.731 & -3.031 & -2.427 \\
          & 3, 0 & -1.263 & -0.985 & -1.285 & -2.207 \\
          & 4, 0 & -2.188 & -1.886 & -2.186 & -1.919 \\
          & 5, 0 & -0.008 & -0.112 & -0.412 & -2.019 \\
          & 6, 0 & -2.402 & -2.217 & -2.517 & -2.490 \\ \hline
1 1 1 4   & 1, 0 &  1.977 &  1.894 &  1.894 & -0.130 \\
          & 3, 0 &  1.322 &  1.005 &  1.005 & -0.172 \\
          & 5, 0 &  1.307 &  0.901 &  0.901 &  0.042 \\ \hline
1 1 4 4   & 1, 0 &  2.080 &  1.071 &  1.071 & -0.061 \\
          & 3, 0 &  1.260 &  0.517 &  0.517 &  0.004 \\
          & 5, 0 &  0.596 &  0.170 &  0.170 &  0.047 \\ \hline
\end{tabular}
\end{center}
\end{table}

  The modification of the centroids of interaction matrix elements
 in the KB3 interaction is discussed in terms of the monopole
 Hamiltonian in Refs. \cite{Caurier,Duflo}. That discussion is
 intimately related to the roles of the $V_{\pi \nu}^{\tau=0}$ force.
 We have shown in the previous papers \cite{Kaneko3,Kaneko4}
 that the $V_{\pi \nu}^{\tau=0}$ force plays an essential role for
 reproducing the binding energy and symmetry energy in a very wide
 range of nuclei.  The $V_{\pi \nu}^{\tau=0}$ force which has a simple
 and definite form (see Eq. (\ref{eq:8}) or (\ref{eq:9})) can play the
 part for Zuker's monopole Hamiltonian.  According to Dufour and Zuker
 \cite{Dufour}, the residual interactions after extracting the monopole
 Hamiltonian or the $V_{\pi \nu}^{\tau=0}$ force must resemble the
 $P_0+P_2+QQ$ force.
 
   The $J$-independent isoscalar $p$-$n$ force  $V_{\pi \nu}^{\tau=0}$
 gives the average value -2.23 MeV to the diagonal matrix elements
 $V(abJ,T=0)$ (see Table \ref{table10}).  The absolute values of
 $V(f_{7/2} p_{3/2} J, T=0)$ of our interaction FEI are larger than
 those of the realistic effective interactions.
 The FPD6 and KB3 interactions have very large matrix elements
 $V(f_{7/2} f_{5/2} J, T=0)$, to which the matrix elements of FEI are
 comparable.  The centroid of $V(f_{7/2} f_{5/2} J, T=0)$ is -1.846
 for FPD6, -1.934 for KB3 and -2.241 for FEI.  Seeing the large
 diagonal matrix elements, one naturally hesitates to omit the
 $f_{5/2}$ orbit from the model space and hence deals with the full
 $fp$ shell space.  Why the truncated space
 $(f_{7/2}, p_{3/2}, p_{1/2})$ works well for the
 $P_0 + P_2 + QQ + V_{\pi \nu}^{\tau=0}$ interaction?  The secret is
 in a special work of $V_{\pi \nu}^{\tau=0}$.  The $J$-independent
 $p$-$n$ force $V_{\pi \nu}^{\tau=0}$, which is a function of only the
 number and isospin of valence nucleons in Eq. (\ref{eq:8}), is
 independent of the model space in fact.  The average contribution of
 $V(f_{7/2} f_{5/2} J, T=0)$ is equivalently taken into account in our
 model.  The residual interactions excluding $V_{\pi \nu}^{\tau=0}$
 are, of course, desired to be taken up.  The inclusion of the
 $f_{5/2}$ orbit will improve wave functions and $B(E2)$ as seen
 in the previous subsection.  There are large off-diagonal matrix
 elements
 $\langle f_{7/2} f_{7/2} J, T=0 |V| f_{7/2} f_{5/2} J, T=0 \rangle$
 and
 $\langle f_{7/2} f_{7/2} J, T=0 |V| f_{5/2} f_{5/2} J, T=0 \rangle$
 in the realistic effective interactions.  Their effects on the
 energies of low-lying states, however, may be secondary as guessed
 from our results.
 
   We have seen the defects of the present model on non-collective
 states and also the insufficiency of the isovector interactions.
 The present interaction remains room for improvement, which could be
 made by comparing with the realistic effective interactions.
 Another way may be to add hexadecapole and hexadecapole-hexadecapole
 forces (octupole and octupole-octupole forces for negative-parity
 states) as an extension of the $P_0 + P_2 + QQ$ force.

\subsection{Properties of the yrast bands in $A$=46, 48
      and 50 nuclei}

\begin{figure}[b]
\begin{center}
    \epsfig{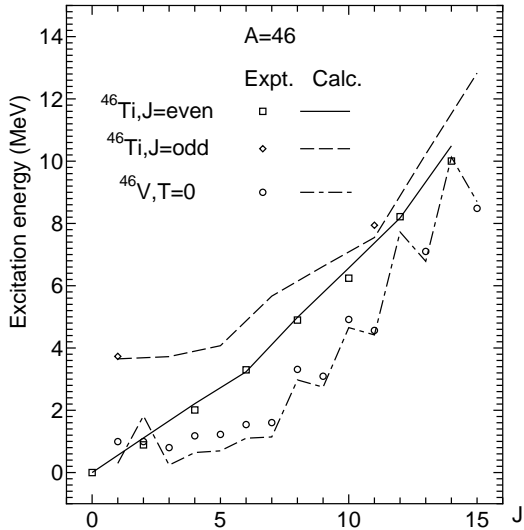}
\caption{Spin-energy relation in the yrast bands of $^{46}$Ti and
         $^{46}$V.}
\label{fig11}
\end{center}
\end{figure}

\begin{figure}[t]
\begin{center}
    \epsfig{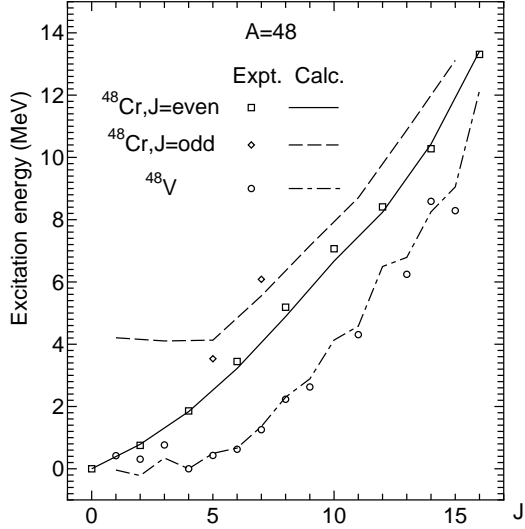}
\caption{Spin-energy relation in the yrast bands of $^{48}$Cr and 
         $^{48}$V.}
\label{fig12}
\end{center}
\end{figure}

\begin{figure}[h]
\begin{center}
    \epsfig{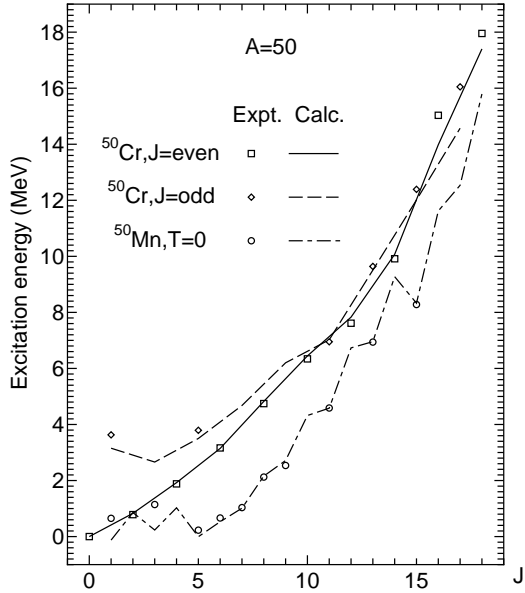}
\caption{Spin-energy relation in the yrast bands of $^{50}$Cr and 
         $^{50}$Mn.}
\label{fig13}
\end{center}
\end{figure}

\begin{figure}[t]
\begin{center}
    \epsfig{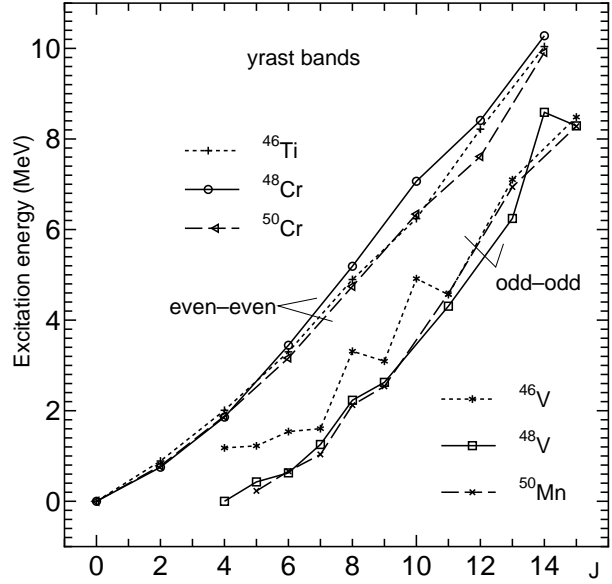}
\caption{Similarity between the observed yrast bands in $A$=46, 48
     and 50 nuclei.}
\label{fig14}
\end{center}
\end{figure}

  In order to look at the structure of the yrast bands in $A$=46, 48
 and 50 nuclei from a different angle, we illustrate the spin-energy
 relation in Figs. \ref{fig11}, \ref{fig12} and \ref{fig13}. We clearly
 see the good applicability of the $P_0+P_2+QQ+V^{\tau =0}_{\pi \nu}$
 interaction to the collective yrast states in these nuclei.
 The rotational behavior and backbending in the ground-state bands
 in even-even nuclei are well reproduced.  It is interesting that
 in $^{50}$Cr (Fig. \ref{fig13}) the $J$=odd line touches the
 ground-state band with $J$=even near $11_1^+$ where the backbending
 occurs and the values of $B(E2:J \rightarrow 11_1^+)$ become large
 (Table \ref{table6}). The staggering gait depending on odd spin and
 even spin observed in the $T=0$ yrast bands of the odd-odd nuclei
 is almost traced by our model.  We have already indicated that the
 magnitudes of staggering are not the same in the cross-conjugate
 nuclei (within the $f_{7/2}$ space) $^{46}$V and $^{50}$Mn.
 Figure  \ref{fig14} shows that the $T=0$ yrast bands of $^{46}$V and
 $^{50}$Mn do not resemble at low spin but resemble at $J$=odd high
 spin. This suggests the dominant contribution of the $f_{7/2}$ orbit
 in the high-spin states of $^{46}$V and $^{50}$Mn.

  Figures \ref{fig11}, \ref{fig12} and \ref{fig13} suggest that there
 are similarities between the ground-states bands of the even-even
 nuclei $^{46}$Ti, $^{48}$Cr and $^{50}$Cr, and between the yrast bands
 of the odd-odd nuclei $^{46}$V, $^{48}$V and $^{50}$Mn.
 In Fig. \ref{fig14}, the observed yrast bands in these nuclei are
 compared in a sheet of drawing.  The slopes at the beginning of the
 bands resemble in the even-even nuclei and also in the odd-odd nuclei
 $^{48}$V and $^{50}$Mn.  This implies a similar mechanism of
 excitation in low-lying collective states of the even-even or odd-odd
 systems, which is considered to be the nuclear rotation
 \cite{Caurier2,Martinez3,Terasaki,Tanaka,Hara}.

 The $N=Z$ odd-odd nuclei $^{46}$V and $^{50}$Mn have $T=1$ and $T=0$
 bands as the lowest and second lowest bands. (Our Hamiltonian gives
 the same energies to the $T=1$ states of $^{46}$Ti and $^{46}$V
 ($^{50}$Cr and $^{50}$Mn). The $T=1$ band observed in $^{46}$V or
 $^{50}$Mn, actually, corresponds well to that observed in $^{46}$Ti
 or $^{50}$Cr.)  Recently, a band crossing which is different from
 the ordinary one caused by the spin alignment of a high $j$ neutron
 or proton pair was observed in a heavy $N=Z$ nucleus
 $^{74}$Rb \cite{Rudolph2}.  The crossing frequency which corresponds
 to $J=5$ or $J=7$ is smaller than that in the ordinary band crossing.
 This phenomenon can be interpreted to be the crossing of the $T=1$
 and $T=0$ bands \cite{Rudolph2,Dean,Kaneko}. Figures \ref{fig11} and
 \ref{fig13} show a similar behavior in $^{46}$V and $^{50}$Mn,
 where the $T=0$ band appears to cross the $T=1$ band near $J=2$.
 We can suppose a similar situation for the $N=Z$ odd-odd nuclei
 both in the beginning and middle regions of $fpg$ shell.
 It is interesting that the crossing frequency in the former region
 is smaller than that in the latter where nuclei are considered to be
 certainly deformed.
 
 The backbending phenomena in $^{48}$Cr and $^{50}$Cr have been
 studied by the cranked Hartree-Fock-Bogoliubov method
 \cite{Caurier2,Martinez3,Terasaki,Tanaka}.  A typical interaction for
 studying the rotation of deformed nucleus by the cranking approach is
 just the $P_0+QQ$ force which is the main part of our residual
 interactions excluding the $V^{\tau =0}_{\pi \nu}$ force.
 It was shown by the projected quasi-particle shell model in the
 Nilsson base \cite{Hara2} that the $P_0 + P_2 + QQ$ force
 successfully describes not only the rotational band in medium-heavy
 nuclei \cite{Hara1} but also the backbending in $^{48}$Cr \cite{Hara}.
 The study of Ref. \cite{Hara} indicates that the $P_0 + P_2 + QQ$
 force is comparable to the realistic effective interaction KB3
 in reproducing the rotational band of $^{48}$Cr. It should be noticed
 here that the $P_0 + P_2 + QQ$ force without the $p$-$n$ force
 $V^{\tau =0}_{\pi \nu}$ cannot well reproduce the binding energy,
 and the $p$-$n$ interactions as well as like-nucleon interactions
 are taken into account in our $P_0 + P_2$ force.
 The $P_0+P_2+QQ+V^{\tau =0}_{\pi \nu}$ interaction is, therefore,
 a very suitable interaction to investigate the rotational band and
 its backbending in the $N \approx Z$ $fp$ shell region,
 in parallel with the shell model approach.

   One of the key words to understand the excitation mechanism
 at the beginning of the $fp$ shell nuclei seems to be $\alpha$-like
 four-nucleon correlations \cite{Ohkubo}. The $^{44}$Ti nucleus is
 an $\alpha$-like correlated state outside the $^{40}$Ca core.
 The $^{48}$Cr nucleus, in zero order approximation, is described
 as a correlated state of two $\alpha$-like clusters (quartets)
 $\sum_I \psi_I (\alpha^\dagger_{I,T=0})^2 |^{40} \mbox{Ca}\rangle $
 where $\alpha^\dagger_{IM,T=0}$ being the lowest-energy Tamm-Dancoff
 modes of four nucleons with spin $IM$ and $T=0$ \cite{Hasegawa3}.
 The approximate description gives the ground-state energy -32.05 MeV
 against the exact energy -32.38 MeV in the present model. Within the
 single $j=f_{7/2}$ shell model, the states of $A=4n+2$ nuclei are
 roughly approximated as follows \cite{Hasegawa2,Hasegawa4}:
\begin{eqnarray}
 |^{46}\mbox{Ti}:J^+_1 \rangle & \approx & \frac{1}{\sqrt{C_1}}
   A^{\dagger}_{0011}(f_{7/2}f_{7/2})
   |^{44}\mbox{Ti}:J^+_1 \rangle \ \ \ (J=0,2), \nonumber \\
 |^{50}\mbox{Cr}:J^+_1 \rangle & \approx & \frac{1}{\sqrt{C_2}}
   A^{\dagger}_{0011}(f_{7/2}f_{7/2})
   |^{48}\mbox{Cr}:J^+_1 \rangle  \ \ \ (J=0,2),   \label{eq:16}
\end{eqnarray}
where $C_1$ and $C_2$ are normalization constants.  The overlaps of
 the approximate states (\ref{eq:16}) with the exact states are more
 than 0.98 for $J=0$ and are more than 0.90 for $J=2$.  In other words,
 the excitation $0^+_1 \rightarrow 2^+_1$ in $^{46}$Ti and $^{50}$Cr
 resembles the excitation $0^+_1 \rightarrow 2^+_1$ in $^{44}$Ti and
 $^{48}$Cr, respectively.  The excitation $0^+_1 \rightarrow 2^+_1$
 in $^{44}$Ti is the change of four nucleon structure
 $\alpha^\dagger_{I=0,T=0} \rightarrow \alpha^\dagger_{I=2,T=0}$.
 A similar change of structure is possibly dominant in the excitation
 $0^+_1 \rightarrow 2^+_1$ in $^{48}$Cr.   We can suppose a common
 excitation mechanism induced by the $\alpha$-like four-nucleon
 correlations in $^{46}$Ti, $^{48}$Cr and $^{50}$Cr.  We have already
 discussed the intimate relation of our interaction
 $P_0+P_2+QQ+V^{\tau =0}_{\pi \nu}$ to the $\alpha$-like four-nucleon
 correlations in Ref. \cite{Hasegawa2}.
 The $P_0+P_2+QQ+V^{\tau =0}_{\pi \nu}$ interaction having strong
 $p$-$n$ interactions may underlie commonly in the nuclear rotation
 and $\alpha$-like four-nucleon correlations in the $fp$ shell nuclei.

\section{Concluding remarks}

  We have applied a functional effective interaction extended from the
 pairing plus $QQ$ force by adding the $J$-independent isoscalar
 $p$-$n$ force $V^{\tau =0}_{\pi \nu}$ and quadrupole pairing force
 to $^{46}$Ti, $^{46}$V, $^{48}$V, $^{48}$Cr, $^{50}$Cr and
 $^{50}$Mn.  The exact shell model calculations in the truncated model
 space ($f_{7/2},p_{3/2},p_{1/2}$) demonstrate the usefulness of the
 interaction for the yrast states in these nuclei with $n_p \ge 2$ and
 $n_n \ge 2$.  The model reproduces well the experimental binding
 energies, energy levels of the yrast states and $B(E2)$ between them.
 We have also analyzed the relationship between our interaction and
 the realistic effective interactions KB3 and FPD6.  The analysis
 clarified the reason why the truncated model space works well
 especially for our interaction. This work as well as the previous
 ones \cite{Hasegawa,Kaneko3,Kaneko4} supports that an important part
 of nucleon-nucleon interactions can be written as
 $V^{\tau =0}_{\pi \nu}$.  The foundation of the $p$-$n$ force
 $V^{\tau =0}_{\pi \nu}$ could be discussed in the framework of the
 HF theory.
 
   The good reproduction of the yrast states made it possible to
 discuss their structure in the $A=$46, 48 and 50 nuclei.  We have
 given some predictions about the energy levels and characteristic
 variations of $B(E2)$ in the yrast bands, in these nuclei.
 The extended $P_0+QQ$ interaction is excel in describing the
 collective nature of nuclei, and is expected to be most suitable
 for studying the rotational properties of the yrast bands.
 The backbending phenomena in $^{48}$Cr and $^{50}$Cr, which are well
 described in terms of the shell model with our interaction, can be
 investigated by a cranking model with the same interaction.
 The $P_0+P_2+QQ+V^{\tau =0}_{\pi \nu}$ interaction composed of
 typical forms of forces is also useful for the study of competition
 between the $T=0$ and $T=1$ $p$-$n$ pairing and like-nucleon pairing
 which is one of current topics.
 There is a sign that the $P_0+P_2+QQ+V^{\tau =0}_{\pi \nu}$
 interaction has an affinity for the $\alpha$-like four-nucleon
 correlations important in $N \approx Z$ nuclei.
 
   The success of the extended $P_0+QQ$ interaction by means of the
 exact shell model gives strong evidence that the extended picture of
 the pairing plus $QQ$ force model holds in lighter nuclei. This
 supports the discussion by the projected quasi-particle shell model
 \cite{Hara1,Hara2,Hara} that the $P_0+P_2+QQ$ interaction describes
 well the rotational states not only in medium-heavy nuclei but also
 in the $fp$ shell nucleus $^{48}$Ca. We can probably say that the
 success of the pairing plus $QQ$ force model in heavier nuclei with
 $N>Z$ did not depend on approximate treatments made there. Our
 calculations in this paper and others \cite{Hasegawa,Kaneko3,Kaneko4}
 clarified the essential role of the $p$-$n$ force
 $V^{\tau =0}_{\pi \nu}$ in the binding energy.
 If $V^{\tau =0}_{\pi \nu}$ is added to the treatment of the pairing
 plus $QQ$ plus quadrupole force model in $N>Z$ nuclei, binding
 energies will be well reproduced too.  
  
   The present model, however, is not sufficiently good for
 non-collective states except the yrast states and for nuclei with
 $n_p \leq 1$ or $n_n \leq 1$. One cause may be attributed to the
 absence of the $f_{5/2}$ orbit in the present calculations and
 another to the insufficiency of the isovector interactions which take
 action in like-nucleon systems.  The present interaction remains room
 for improvement.  One way to improve it is to add hexadecapole and
 hexadecapole-hexadecapole forces (octupole and octupole-octupole
 forces for negative-parity states) to the $P_0+P_2+QQ$ force.



\end{document}